\documentclass[runningheads]{llncs}
\usepackage[T1]{fontenc}
\usepackage{graphicx}
\usepackage{color}  
\usepackage{amsmath,amssymb} 
\usepackage{comment} 
\newtheorem{assumption}{Assumption}
\usepackage{algorithmic}
\usepackage{algorithm}

\newcommand{\argmin}{\mathop{\rm arg~min}\limits}
\usepackage{listings}
\usepackage{subcaption}

\begin{document}
\title{Algebraically Observable Physics-Informed Neural Network and 
its Application to Epidemiological Modelling}
%
%
\author{Mizuka Komatsu\inst{1}\orcidID{0000-0002-8482-524X} 
}
\authorrunning{M. Komatsu}
%
\institute{
Graduate School of System Informatics, Kobe University, Kobe, Hyogo, Japan\\
\email{m-komatsu@bear.kobe-u.ac.jp}
}

\maketitle              
%

%
\begin{abstract}
A physics-informed neural network (PINN) is a deep learning framework that integrates the governing equations underlying data into a loss function. 
In this study, we consider the problem of estimating state variables and parameters in epidemiological models governed by ordinary differential equations using PINNs. In practice, not all trajectory data corresponding to the population described by models can be measured. Learning PINNs to estimate the unmeasured state variables and epidemiological parameters using partial measurements is challenging. 

Accordingly, we introduce the concept of algebraic observability of the state variables. Specifically, we propose augmenting the unmeasured data based on algebraic observability analysis. The validity of the proposed method is demonstrated through numerical experiments under three scenarios in the context of epidemiological modelling. 
Specifically, given noisy and partial measurements, the accuracy of unmeasured states and parameter estimation of the proposed method is shown to be higher than that of the conventional methods. 
The proposed method is also shown to be effective in practical scenarios, such as when the data corresponding to certain variables cannot be reconstructed from the measurements.

\end{abstract}
\section{Introduction}\label{sec:intro}
This study focuses on estimating the state and parameter variables of epidemiological models. Throughout this work, epidemiological models refer to compartmental models described by differential equations, such as the Susceptible–-Infectious–-Recovered (SIR) and Susceptible–-Exposed–-Infectious-–Recovered (SEIR) models \cite{kuniya,inaba,massonis}. The state variables in such epidemiological models represent the subpopulation or its proportion divided by the infection status; the parameters represent quantities such as infection and recovery rates. The estimation of these model variables is particularly important as it plays a key role in predicting outbreaks and evaluating disease control measures \cite{kuniya,pnas_covid,Rinaldi}. 

We explore the use of Physics-Informed Neural Networks (PINNs) \cite{pinn}, which have recently attracted attention as a method for state and parameter estimation in systems governed by differential equations. PINNs are a class of neural networks that embed physical laws or governing equations into their architecture, enabling them to incorporate prior knowledge of the system behaviour.
Owing to the expressiveness of neural networks, PINNs have shown promising results in estimating state variables and parameters when complete measurements are available \cite{SciML-PINN,pinn,Epi-DNNs,mpinn}.
However, in cases of partial observations, that is, when some state variables are not measured, PINNs may fail to accurately estimate both the states and parameters \cite{mpinn}. This situation arises frequently in epidemiological modelling because certain infection statuses are often difficult or even impossible to measure in practice \cite{mpinn,massonis}. 

In addition to PINNs, alternative approaches for state and parameter estimation in nonlinear state--space models include methods based on Kalman \cite{ekf_covid,ifac_kf} and particle filters \cite{comp_pf}. Regarding Kalman filters, it is common to assume that the parameters are quasistationary and include them in the state vector by applying an extended Kalman filter (EKF) to the resulting augmented system. 
In general, the EKF approximates the dynamics of a model augmented with parameters, which introduces a discrepancy between the governing equations and the equations used for state and parameter estimation. 
Therefore, if parameter values can be accurately estimated, PINNs, which possess universal approximation capabilities, are expected to yield more accurate predictions than Kalman filter-based methods. In addition, although there is ongoing research on EKF under partial observation \cite{ifac_kf}, such methods remain generic in nature. In particular, the characteristics of the governing equations, such as the relationship between the observed and unobserved variables, are rarely considered. The particle-filter based methods \cite{comp_pf,epi_pmcmc,epo_smc2} are usually more general than the Kalman filter-based methods. These approaches utilise particles to represent the distributions of parameters and/or states estimates. This enables greater expressive power in representing state variables compared to methods based on Kalman filters. However, it has been pointed out that strong dependencies between parameters and state variables can hinder the effectiveness of methods based on particle filters \cite{comp_pf}. 
Similar to Kalman filter-based methods, the particle-filter based methods often do not consider equation-specific properties. 
See Section \ref{sec:conclusion} for further discussion. 

Based on this background, we focus on the problem of state and parameter estimation in epidemiological models under partial observation settings and propose a modified version of PINN to improve the accuracy of the estimation. 
The proposed method, called algebraically observable PINN, is based on algebraic observability analysis \cite{meshkat}. This enables the estimation to take advantage of the relationship between observed and unobserved variables. 
In general, a state variable can be observed if it is constructed from the data \cite{kalman}. In principle, if a state variable is shown to be observable through algebraic observability analysis, the variable can be explicitly constructed from the measured data. Building on this idea, we introduce an algebraic analysis to augment the data corresponding to unmeasured variables, thereby aiming to improve the accuracy of the estimation.

The remainder of this paper is organised as follows. 
Section \ref{sec:background} summarises the fundamental aspects of PINNs. The problem addressed in this study is formulated in Section \ref{subsec:partial}. Section \ref{subsec:illustration} presents a specific example of the problem and highlights the limitations of existing methods.
The proposed method, which is based on algebraic observability analysis, is explained in Section \ ref{sec: proposed}. We refer to this method as algebraically observable PINN. 
In Section \ref{sec:exp}, we demonstrate the validity of the proposed method through numerical experiments, considering three scenarios for epidemiological modelling. Finally, the paper concludes with a discussion and conclusions. 

\section{Background: Physics-Informed Neural Networks}\label{sec:background}
In general, two major problems are associated with PINNs \cite{pinn}. The first involves solving the partial differential equations (PDEs) with known parameters and initial and boundary conditions. 
In this problem, which is known as the forward problem of PINNs, the solution of PDEs is parameterised by a neural network.
The second problem is the inverse problem: given the measurements from the solution of PDEs, unknown parameters related to PDEs are estimated while approximating the solutions using PINNs. 
In the following sections, these problems are explained, based on the settings considered in this study. 
In Section \ref{subsec:partial}, the limitations of conventional approaches are clarified.
\subsection{Forward Problems}\label{subsection:pinn_forward}
Throughout this study, we consider the governing equations described by a system of ordinary differential equations (ODEs) defined in the time interval $[0, T]$ as follows:
\begin{align}
\dot{x} &= f(x, u; \theta), \label{eq:model}    
\end{align}
where $f:\mathbb{R}^{N+L} \rightarrow \mathbb{R}^N$.
$x = {(x_1, \ldots,x_N)}^\top\in \mathbb{R}^N$ denotes the state variable vector of the system under consideration; $u = {(u_1, \ldots, u_L)}^\top \in \mathbb{R}^L$ denotes the input vector; $\theta = {(\theta_1, \ldots, \theta_L)}^\top \in \mathbb{R}^n$ denotes the parameter vector. 
We assume that $u$ is a known and sufficiently smooth function vector defined over the considered period. 
Throughout this study, these parameters are maintained constant.
We also assume that each component $f_i$ of the vector field $f = {(f_1, \ldots, f_N)}^\top$ is a polynomial of $x, u$ with coefficients in $\mathbb{Q}(\theta)$, $f_i \in \mathbb{Q}(\theta)[x, u]$ for $i = 1,\ldots, N$. 
In this paper, the variables with dots represent their derivatives with respect to $t$. 
However, depending on the context, we denote the $s$th order derivative ${\mathrm{d}^{s} x}/{\mathrm{d} t^{s}}$ as $x^{(s)}$ rather than using dots. 
\begin{example}[SAIRD model \cite{saird_u}]\label{example:saird}
The SAIRD model \cite{saird_u} is an epidemiological model that is defined as follows:
\begin{align}
\begin{split}
\dot{S} &= -\beta u S I- \xi u S A,\\
\dot{A} &= \beta u S I+ \xi u S A - \kappa A,\\
\dot{I} &= \kappa A-(\gamma+\delta)I, \\
\dot{R} &= \gamma I,\\
\dot{D} &= \delta I.
\end{split}\label{eq:saird}
\end{align}
Similar to other epidemiological models, the SAIRD model assumes that a population is divided into subgroups. 
$S$, $A$, $I$, $R$, and $D$ denote the population ratios of each subgroup, that is, susceptible, asymptomatic, infectious, recovered, and dead respectively. 
The transitions between the subgroups are described in the system of ODEs in \eqref{eq:saird}, where ${(S, A, I, R, D)}^\top \in \mathbb{R}^5$ is a state vector that depends on time $t$. 
$(\beta, \xi, \kappa, \gamma, \delta) \in \mathbb{R}^5$ correspond to the parameter vectors representing the infection, transmission, progression, recovery, and rate of dead, respectively. 
$u: [0, T] \rightarrow \mathbb{R}$ is a function of $t$ that represents the change in infection and transmission rates with respect to time. In \cite{saird_u}, $u$ is modelled as $u(t) = \exp(-kt)$. In this case, if the value of $k \in \mathbb{R}$ is known, $u(t)$ can be regarded as a known input function in \eqref{eq:saird}. The right side of \eqref{eq:saird} can be regarded as a polynomial in $\mathbb{Q}(\beta, \xi, \kappa, \gamma, \delta)[S, A, I, R, D, u]$.  
\end{example}

In the forward problem, we assume the following initial conditions: 
\begin{align*}
    x(0) = \bar{x}_0 = {(\bar{x}_1(0), \ldots, \bar{x}_N(0))}^\top\in \mathbb{R}^N
\end{align*}
and parameter vector $\theta$ is known and fixed. 
Given the evaluation time points, $\mathcal{T} = \left\{ t_d \in [0, T] \mid d = 1, \ldots, D\right\}$
and the evaluation of $u$ at time point $\mathcal{T}$, 
a neural network 
\begin{align*}
    x_\mathrm{nn}(t; \theta_\mathrm{nn}) = {(x_{\mathrm{nn}, 1}(t; \theta_\mathrm{nn}), \ldots, x_{\mathrm{nn}, N}(t; \theta_\mathrm{nn}))}^\top: \mathbb{R} \rightarrow \mathbb{R}^N
\end{align*} is learned to approximate the solution of \eqref{eq:model}. 
where $\theta_\mathrm{nn} \in \mathbb{R}^{n'}$ denotes a network parameter vector. 
To incorporate the governing equations \eqref{eq:model} and initial conditions in the network as prior knowledge, $x_\mathrm{nn}(t; \theta_\mathrm{nn})$ is learned by minimising the following loss function:
\begin{align}
\begin{split}
L_0(\theta_\mathrm{nn}; \bar{x}_0, u)
= \lambda_{\mathrm{eq}}L_{\mathrm{eq}}(\theta_\mathrm{nn}; u)  + \lambda_{\mathrm{init}}L_{\mathrm{init}}(\theta_\mathrm{nn}; \bar{x}_0), \quad\lambda_\mathrm{eq}, \lambda_\mathrm{init} \in \mathbb{R}
\end{split}\label{eq:loss_forward}
\end{align}
where $L_{\mathrm{eq}}(\theta_\mathrm{nn}; u)$ and $L_{\mathrm{init}}(\theta_\mathrm{nn}; \bar{x}_0)$ are the deviations of the governing equations and initial conditions, respectively. 
Conventionally, $L_{\mathrm{eq}}(\theta_\mathrm{nn})$ and $L_{\mathrm{init}}(\theta_\mathrm{nn}; \bar{x}_0)$ are defined as 
\begin{align}
L_{\mathrm{eq}}(\theta_\mathrm{nn}; u) &= \sum_{i\in \mathcal{N}} 
\left(\frac{1}{|\mathcal{T}|}\sum_{d=1}^{|\mathcal{T}|} {(\dot{x}_{\mathrm{nn},i}(t_d;\theta_\mathrm{nn}) - f_i(x_\mathrm{nn}(t_d;\theta_\mathrm{nn}), u(t_d); \theta))}^2\right),\label{eq:leq}\\
L_{\mathrm{init}}(\theta_\mathrm{nn}; \bar{x}_0) &= \sum_{i\in \mathcal{N}}{({x}_{\mathrm{nn},i}(0;\theta_\mathrm{nn}) - \bar{x}_i(0))}^2, \label{eq:linit}
\end{align}
where $\mathcal{N} = \left\{1, \ldots, N\right\}$.
$\dot{x}_{\mathrm{nn}}(t; \theta_\mathrm{nn}) = {(\dot{x}_{\mathrm{nn},1}(t; \theta_\mathrm{nn}), \ldots, \dot{x}_{\mathrm{nn},N}(t; \theta_\mathrm{nn}))}^\top$ denotes the derivative of ${x}_{\mathrm{nn}}(t; \theta_\mathrm{nn}) = {({x}_{\mathrm{nn},1}(t; \theta_\mathrm{nn}), \ldots, {x}_{\mathrm{nn},N}(t; \theta_\mathrm{nn}))}^\top$, which is computed by automatic differentiation, and thus, is exact.
In the forward problem of PINN, \eqref{eq:loss_forward} is minimised with respect to $\theta_\mathrm{nn}$, and thus, $x_\mathrm{nn}(t; \theta_\mathrm{nn})$ that satisfies \eqref{eq:model}, and the initial conditions are obtained to the extent possible. Compared to numerical methods for solving ODEs, PINNs exploit neural networks as universal function approximators. As will be discussed later, the network architectures for inverse problems are essentially the same as those for forward problems; hence, the expressibility of neural networks is also utilised in inverse problems.

\subsection{Inverse problems}\label{subsec:pinn_inverse}
In this section, we describe the conventional setup of inverse problems for PINNs.
In addition to the initial conditions $x(0) = \bar{x}_0$ and the evaluations of $u$ at time points in $\mathcal{T}$, we assume that the evaluations of the solution to \eqref{eq:model} at the time points in $\mathcal{T}$ are provided as measurement data. 
Specifically, the measured data $\bar{x}^\mathcal{T}$ are expressed as follows:
\begin{align}
\left\{ \bar{x}(t_d) := x(t_d; \theta_*, \bar{x}_0, u) \mid t_d \in \mathcal{T} \right\},
\end{align}
where the solution to \eqref{eq:model}, given the fixed parameter values $\theta = \theta_*$, $x_0$ and $u$ is denoted as $x(t; \theta_*, \bar{x}_0, u)$. 
\begin{remark}
In Sections \ref{sec:background} and \ref{sec:proposed}, we assume noise-free measurements. 
In this study, this assumption is adopted to integrate PINNs with algebraic approaches. 
However, PINNs can also be applied to noisy datasets. For details regarding the robustness of PINNs, see \cite{pinn,SciML-PINN}. The robustness of the proposed method is discussed later in Section \ref{sec:exp}. 
\end{remark}
In the inverse problems of PINNs, $\theta_*$ is assumed to be unknown and estimated from the measured data. 
As $\theta$ is assigned specific meanings in relation to the governing equations in \eqref{eq:model}, estimating these values is meaningful. 
For example, in epidemiological models, infection rates are important for predicting infection trends. 
Under the framework of PINNs, the value of $\theta$ is estimated while learning $\theta_\mathrm{nn}$ such that $x_\mathrm{nn}(t; \theta_\mathrm{nn})$ fits the measured data and satisfies \eqref{eq:model} and the initial conditions. 
Thus, given the constants $\lambda_\mathrm{eq}, \lambda_\mathrm{init}, \lambda_\mathrm{data} \in \mathbb{R}$, the loss function 
\begin{align}
\begin{split}
&L(\theta_\mathrm{nn}, \theta; \bar{x}_0, u, \bar{x}^\mathcal{T})\\
&= \lambda_{\mathrm{eq}}L_{\mathrm{eq}}(\theta_\mathrm{nn}, \theta; u)  
+ \lambda_{\mathrm{init}}L_{\mathrm{init}}(\theta_\mathrm{nn}; \bar{x}_0) 
+ \lambda_{\mathrm{data}}L_{\mathrm{data}}(\theta_\mathrm{nn}; \bar{x}^\mathcal{T}) 
\end{split}\label{eq:loss_inverse}
\end{align}
is minimised with respect to $(\theta_\mathrm{nn}, \theta) \in \mathbb{R}^{n'}\times \mathbb{R}^n$. 
Similar to the forward problem, $L_{\mathrm{eq}}(\theta_\mathrm{nn}, \theta; u)$ and $L_{\mathrm{init}}(\theta_\mathrm{nn}; \bar{x}_0)$ evaluate deviations in the governing equations and initial conditions, respectively. 
These are defined on the right sides of \eqref{eq:leq} and \eqref{eq:linit}, respectively. Notably, $L_{\mathrm{eq}}(\theta_\mathrm{nn}, \theta; u)$ is regarded as a function of $(\theta_\mathrm{nn}, \theta)$ because an unknown $\theta$ appears in \eqref{eq:model}. Regarding \eqref{eq:loss_inverse}, $L_{\mathrm{data}}(\theta_\mathrm{nn}; \bar{x}^\mathcal{T})$ evaluates the deviation in the measured data and is defined as follows:
\begin{align}
L_{\mathrm{data}}(\theta_\mathrm{nn}; \bar{x}^\mathcal{T}) 
&= \sum_{i\in \mathcal{N}}
\left(\frac{1}{|\mathcal{T}|}\sum_{d=1}^{|\mathcal{T}|}{({x}_{\mathrm{nn},i}(t_d;\theta_\mathrm{nn}) - \bar{x}_i(t_d))}^2\right).\label{eq:ldata}
\end{align}
In the inverse problem of PINNs, \eqref{eq:loss_inverse} is minimised expecting to obtain $(\hat{\theta}_\mathrm{nn}, \hat{\theta})$ such that it holds $x_\mathrm{nn}(\cdot; \hat{\theta}_\mathrm{nn}) \simeq x(\cdot; \theta_*, \bar{x}_0, u)$ and $\theta \simeq \theta_*$. 
Conventionally, \eqref{eq:loss_inverse} is minimised using stochastic gradient descent and its modern variants. 
Thus, using the PINN, state and parameter estimations were performed simultaneously. Furthermore, after learning the PINN, the trained network can be applied to predict target dynamics. 

\subsection{Inverse problems given partial measurements}\label{subsec:partial}
In practice, some state variables may not be measured or may be highly noisy, owing to measurement costs or measurement environments, respectively. In this study, we consider inverse problems with partial measurements as the target problems. In epidemiological models, owing to the limitations of testing capacity, unknown parameters are typically estimated based on partial measurements \cite{massonis}. Therefore, considering PINNs for the given partial measurements is worthwhile \cite{mpinn}. 

In the following, we provide a detailed description of the inverse problems of PINNs given partial measurements along with naive approaches. 
We denote the set of indices of the state variables as $\mathcal{N} = \left\{1, \ldots, N \right\}$ and one of the measured state variables as $\mathcal{M} \subseteq \mathcal{N}$. If all state variables are measured, then it holds that $\mathcal{N} = \mathcal{M}$. 
This corresponds to the case considered in Section \ref{subsec:pinn_inverse}. 
Using the notation $\mathcal{N}$ and $\mathcal{M}$, the set of indices of the unmeasured state variables can be represented as $\mathcal{N} \backslash \mathcal{M}$. 

Given $\mathcal{M}$, we consider the (partial) measurement of the state variables $\bar{x}^\mathcal{T}_\mathcal{M}$ as follows:
\begin{align}
\left\{
\begin{aligned}
\bar{x}_\mathcal{M}(t_d)  
&= {
\left( 
\bar{x}_{\mathrm{idx}(1)}(t_d), 
\cdots, 
\bar{x}_{\mathrm{idx}(|\mathcal{M}|)}(t_d)
\right)}^\top \\
&:= 
{\left( 
x_{\mathrm{idx}(1)}(t_d; \theta_*, \bar{x}_0, u), 
\cdots, 
x_{\mathrm{idx}(|\mathcal{M}|)}(t_d; \theta_*, \bar{x}_0, u)
\right)}^\top 
\end{aligned}
\;\middle|\; t_d \in \mathcal{T} 
\right\},
\label{eq:obs_M}
\end{align}
where $\mathrm{idx}(i)$ denotes the $i$th smallest value in $\mathcal{M}$ and $|\mathcal{M}|$ denotes the size of $\mathcal{M}$. We note that $x^\mathcal{T}_\mathcal{N} = x^{\mathcal{T}}$. 
The naïve choice of loss functions for PINNs given partial measurements $\bar{x}^\mathcal{T}_\mathcal{M}$ is as follows:
\begin{align}
\begin{split}
&L_{\mathcal{M}}(\theta_\mathrm{nn}, \theta; \bar{x}_0, u, \bar{x}^\mathcal{T}_{\mathcal{M}})\\
&= \lambda_{\mathrm{eq}}L_{\mathrm{eq}}(\theta_\mathrm{nn}, \theta; u)  
+ \lambda_{\mathrm{init}}L_{\mathrm{init}}(\theta_\mathrm{nn}; \bar{x}_0) 
+ \lambda_{\mathrm{data}}L_{\mathrm{data}}(\theta_\mathrm{nn}; \bar{x}^\mathcal{T}_\mathcal{M}), 
\end{split}\label{eq:loss_M}
\end{align}
where 
$L_{\mathrm{data}}(\theta_\mathrm{nn}; \bar{x}^\mathcal{T}_\mathcal{M})$ are defined as follows:
\begin{align}
L_{\mathrm{data}}(\theta_\mathrm{nn}; \bar{x}^\mathcal{T}_\mathcal{M}) 
= 
\sum_{i \in \mathcal{M}}
\left(\frac{1}{|\mathcal{T}|}\sum_{d=1}^{|\mathcal{T}|}{({x}_{\mathrm{nn},i}(t_d;\theta_\mathrm{nn}) - \bar{x}_i(t_d))}^2\right).\label{eq:ldataM}
\end{align}
Note that \eqref{eq:ldata} can be considered a special case of \eqref{eq:ldataM}, where $\mathcal{M} = \mathcal{N}$.
If the initial values of the states are fully measured, \eqref{eq:loss_M} can be applied. Partial measurements of the initial conditions 
\begin{align}
\bar{x}_{0, \mathcal{M}} = 
{\left(
\bar{x}_{\mathrm{idx}(1)}(0), \ldots, \bar{x}_{\mathrm{idx}(|\mathcal{M}|)}(0)
\right)}^\top
\end{align}
are given, 
\begin{align}
L_{\mathrm{init}}(\theta_\mathrm{nn}; \bar{x}_{0, \mathcal{M}}) 
&= \sum_{i\in\mathcal{M}}{({x}_{\mathrm{nn},i}(0;\theta_\mathrm{nn}) - \bar{x}_i(0))}^2
\label{eq:linitM}
\end{align}
is introduced instead of \eqref{eq:linit}, which is a special case of \eqref{eq:linitM}, where $\mathcal{M} = \mathcal{N}$. 

The inverse problems of PINNs given partial measurements, that is, $\mathcal{M} \subset \mathcal{N}$, are substantially challenging, as highlighted in a previous study \cite{mpinn}. Owing to the lack of measurements $\bar{x}_i (i \notin \mathcal{M})$, a degree of freedom exists in learning $x_\mathrm{nn}(t; \theta_\mathrm{nn})$ based on \eqref{eq:loss_M}. As mentioned earlier, the loss functions of PINNs are minimised with respect to $\theta_\mathrm{nn}$ and $\theta$ simultaneously. Therefore, such degrees of freedom may lead to failures in learning not only $x_\mathrm{nn}(t; \theta_\mathrm{nn})$ but also $\theta$. 
To clarify this situation, we provide an example of the inverse problem of PINNs given partial measurements and how the naive approach fails in Section \ref{subsec:illustration}.

\subsection{Illustrative example}\label{subsec:illustration}
In this section, an example is provided to highlight the difficulty in solving inverse problems of PINNs given partial measurements.
We consider the following SEIR model \cite{massonis,kuniya}: 
\begin{align}
\begin{split}
\dot{S} = -{\beta}{S}I,\quad
\dot{E} = {\beta}{S}{I}-{\epsilon}{E},\quad
\dot{I} = {\epsilon}{E}-{\gamma}{I}, \quad
\dot{R} = {\gamma}{I}.
\end{split}\label{eq:seir}
\end{align}
The SEIR model assumes that a population is divided into four subgroups.
$S$, $E$, $I$, and $R$ denote the population ratios of each subgroup, that is, susceptible, exposed, infectious, and recovered, respectively. The transitions between the subgroups are described in \eqref{eq:seir}, where $x = {(x_1, x_2, x_3, x_4)}^\top = {(S, E, I, R)}^\top$ is the state vector. 
Three epidemiological parameters are present: $\beta$, $\epsilon$, $\gamma$, which denote the infection, onset, and recovery rates, respectively. 
We assume that the values of $\beta, \gamma$ and the initial conditions are $(\beta, \gamma) = (0.26, 0.1), \bar{x}_0 = (\bar{x}_1(0), \bar{x}_2(0), \bar{x}_3(0), \bar{x}_4(0)) = (0.99, 0.0, 0.01, 0.0)$. 
We consider an inverse problem in which an unknown parameter $\epsilon$ is estimated using the time-series measurements of $I$ generated by \eqref{eq:seir}, given $\epsilon = \epsilon_* = 0.2$. It holds that $\mathcal{N} = \left\{1, \ldots, 4\right\}, \mathcal{M} = \left\{3\right\}$. 
The measured time points for the training dataset were {$|\mathcal{T}_\mathrm{train}| = 100$} equally spaced points in the interval $[0,200]$, and are represented as follows:
\begin{align*}
x^{\mathcal{T}_\mathrm{train}}_\mathcal{M} := 
\left\{
\bar{x}_3(t_d) : =  x_3(t;(\beta, \gamma, \epsilon_*), \bar{x}_0)
\mid  t_d \in {\mathcal{T}_\mathrm{train}}
\right\}\
\end{align*}
where 
${\mathcal{T}_\mathrm{train}} = \left\{(\Delta t)d \mid \Delta t = 200/100, d = 1, \ldots, 100 \right\}$. 
No input function exists for \eqref{eq:seir}. 
The Dormand--Prince method with a time-step size $\Delta t = 200/1000$ is used as the numerical solver. 
For the test dataset, 100 time points over $[0, 200]$ are selected randomly. 

The purpose of this problem is to estimate $\epsilon \simeq \epsilon_*$ and train $x_\mathrm{nn}(t; \theta_\mathrm{nn}) \simeq x(t;(\beta, \gamma, \epsilon_*), \bar{x}_0)$. 
The loss function is introduced in \eqref{eq:loss_M}, where
$L_\mathrm{eq}(\theta_\mathrm{nn}, \theta)$ and $L_\mathrm{data}(\theta_\mathrm{nn}, \theta;  x^\mathcal{T}_\mathcal{M})$ are defined as follows:
\begin{align}
\begin{split}
&L_\mathrm{eq}(\theta_\mathrm{nn}, \theta) \\
&= \frac{1}{|\mathcal{T}_\mathrm{train}|}\sum_{d=1}^{|\mathcal{T}_\mathrm{train}|}{\left(\dot{x}_{\mathrm{nn},1}(t_d;\theta_\mathrm{nn})+ {\beta}{x_{\mathrm{nn}, 1}(t_d;\theta_\mathrm{nn})}x_{\mathrm{nn, 3}}(t_d;\theta_\mathrm{nn})\right)}^2\\
&+\frac{1}{|\mathcal{T}_\mathrm{train}|}\sum_{d=1}^{|\mathcal{T}_\mathrm{train}|}{\left(\dot{x}_{\mathrm{nn},2}(t_d;\theta_\mathrm{nn})-\left( {\beta}{x_{\mathrm{nn}, 1}(t_d;\theta_\mathrm{nn})}{x_{\mathrm{nn}, 3}(t_d;\theta_\mathrm{nn})}-{\epsilon}{x_{\mathrm{nn}, 2}(t_d;\theta_\mathrm{nn})}\right)\right)}^2\\
&+\frac{1}{|\mathcal{T}_\mathrm{train}|}\sum_{d=1}^{|\mathcal{T}_\mathrm{train}|}{\left(\dot{x}_{\mathrm{nn},3}(t_d;\theta_\mathrm{nn})-\left({\epsilon}x_{\mathrm{nn}, 2}(t_d;\theta_\mathrm{nn})-{\gamma}x_{\mathrm{nn},3}(t_d;\theta_\mathrm{nn})\right)\right)}^2\\
&+\frac{1}{|\mathcal{T}_\mathrm{train}|}\sum_{d=1}^{|\mathcal{T}_\mathrm{train}|}{\left(\dot{x}_{\mathrm{nn},4}(t_d;\theta_\mathrm{nn})+{\gamma}x_{\mathrm{nn},3}(t_d;\theta_\mathrm{nn})\right)}^2,
\end{split}\label{eq:seir_leq}\\
&L_{\mathrm{data}}(\theta_\mathrm{nn}, \theta; x^\mathcal{T}_\mathcal{M}) =
\frac{1}{|\mathcal{T}_\mathrm{train}|} \sum_{d=1}^{|\mathcal{T}_\mathrm{train}|} {(x_{\mathrm{nn},3}(t_d;\theta_\mathrm{nn}) - \bar{x}_3(t_d))}^2.
\label{eq:seir_ldata}
\end{align}
$L_\mathrm{init}(\theta_\mathrm{nn}, \theta; \bar{x}_0)$ are defined by \eqref{eq:linit}. The weight constants $(\lambda_\mathrm{eq}, \lambda_\mathrm{init}, \lambda_\mathrm{data})$ are set to $(1, 1, 1)$.
Based on this loss function, a neural network $x_\mathrm{nn}(t; \theta_\mathrm{nn}): \mathbb{R} \rightarrow \mathbb{R}^4$ was trained. 
For the architecture of $x_\mathrm{nn}(t; \theta_\mathrm{nn})$ and the details of the training, see Section \ref{subsection:ex_setting}. 
As a reference, the PINN was also trained based on \eqref{eq:loss_M}, given the full measurements, $\mathcal{N} = \mathcal{M}$.

\begin{figure}
\begin{center}
\includegraphics[width=10.5 cm]{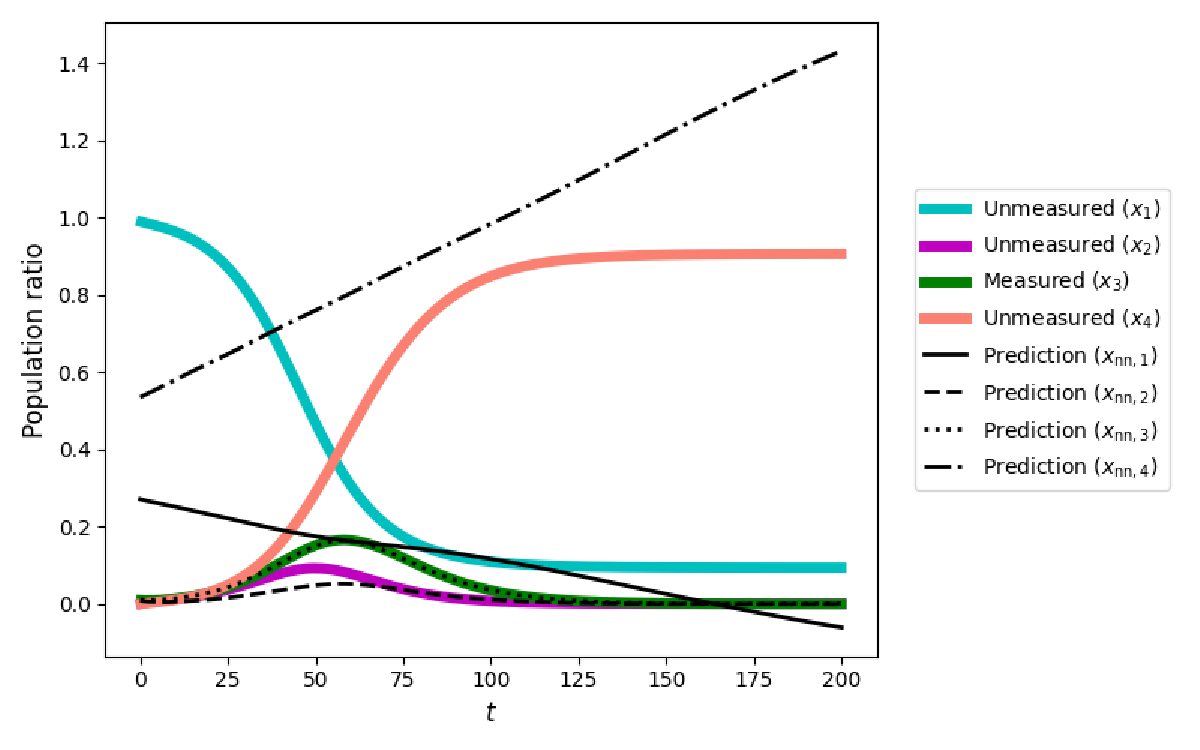}
\caption{Results of state estimation based on the minimisation of \eqref{eq:loss_M} given partial measurements. The coloured and solid lines represent the ground truth, that is, the numerical solution of \eqref{eq:seir} given $\epsilon_*$. The black lines represent the prediction by $x_\mathrm{nn}(t; \theta_\mathrm{nn})$. Note that the measured time-series data are shown as the green solid line.}
\label{fig:results_partial}
\end{center}
\end{figure}

\begin{figure}
\begin{center}
\includegraphics[width=10.5 cm]{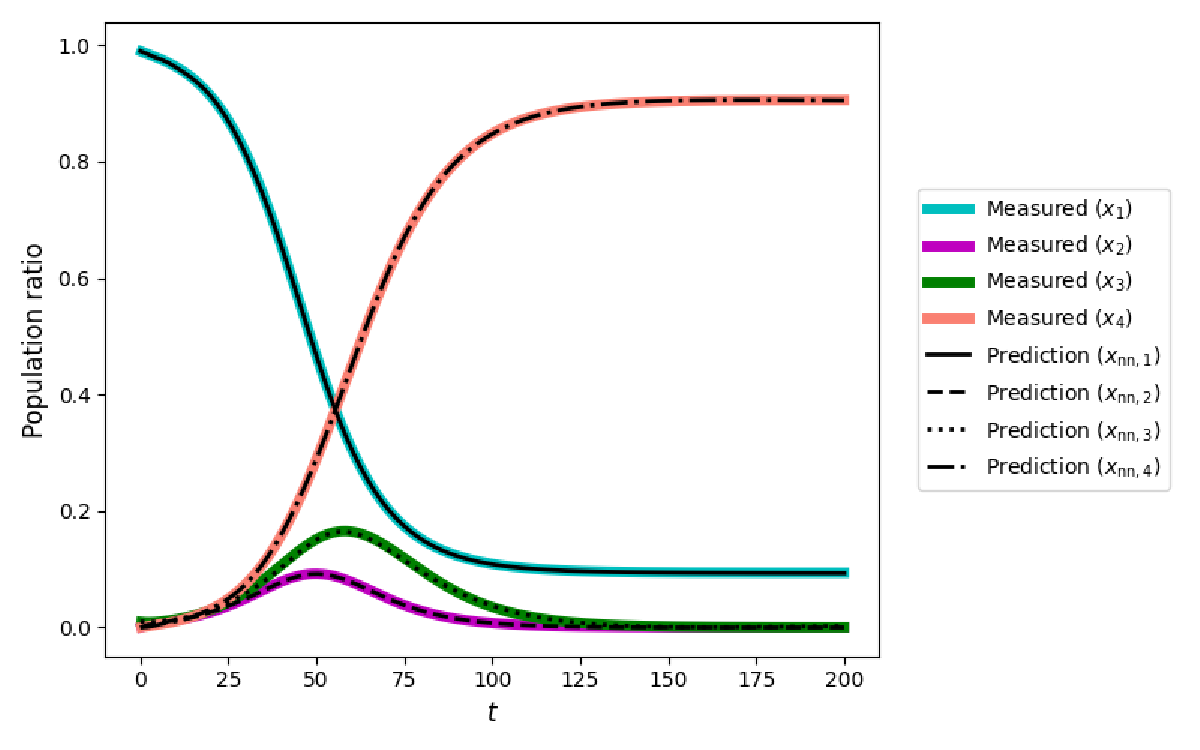}
\caption{Results of state estimation based on the minimisation of \eqref{eq:loss_M} given full measurements. The coloured and solid lines represent the ground truth, that is, the numerical solution of \eqref{eq:seir} given $\epsilon_*$. The black lines represent the prediction by $x_\mathrm{nn}(t; \theta_\mathrm{nn})$.}
\label{fig:results_full}
\end{center}
\end{figure}

Figs. \ref{fig:results_partial} and \ref{fig:results_full} show the results of state estimation given partial and full measurements, respectively. 
The black lines show the prediction by the trained $x_\mathrm{nn}(t; \theta_\mathrm{nn})$ at which point the test loss reaches its minimum values. 
In Fig. \ref{fig:results_partial}, the prediction of unmeasured states $x_{\mathrm{nn},1}, x_{\mathrm{nn},2}, x_{\mathrm{nn},4}$ exhibit minimal agreement with the ground truth. By contrast, the prediction of the measured state $x_{\mathrm{nn},3}$ fits the measured data better than those of the other variables. 
In Fig. \ref{fig:results_full}, all the states fit the ground truth uniformly. For quantitative evaluation, the relative $L_2$ errors for each state variable evaluated using the test data are summarised in Table \ref{tab:relL2_illustrative}. Owing to the lack of measurements corresponding to $x_1 (= S), x_2 (= E), x_4(= R)$, the PINN has one degree of freedom, resulting in difficulty in learning $x_\mathrm{nn}(t; \theta_\mathrm{nn})$.


\begin{table}
\caption{
Relative squared error between the prediction of each state variable and the ground truth evaluated using the test dataset. All values are scaled by $10^{-3}$.
The first row corresponds to the results given partial measurements; $\mathcal{M} = \left\{3\right\}$.
The second row corresponds to the results given full measurements; $\mathcal{M} = \mathcal{N}$.
\label{tab:relL2_illustrative}
}
\begin{center}
\begin{tabular}{|c|c|c|c|c|}
\hline
\textbf{Measurements} & \textbf{$x_1$}	&\textbf{$x_2$} &\textbf{$x_3$}   &\textbf{$x_4$} \\
\hline
Partial
&$724.28$
&$478.54$
&$13.55$
&$523.80$\\
Full 
&$1.63$
&$11.96$
&$26.67$
&$0.96$\\
\hline
\end{tabular}
\end{center}
\end{table}
Given partial measurements, the estimated value of $\epsilon$, denoted by $\hat{\epsilon}$, is $0.242$.
With full measurement, $\hat{\epsilon} = 0.187$.
The relative absolute error $|\epsilon - \epsilon_* |/ |\epsilon_*|$ was $0.21$ for the partial measurements and $0.065$ for the full measurements. Thus, partial measurements make it difficult to appropriately estimate not only the state variables, but also the parameters. 

To clarify this situation, we present an evaluation of \eqref{eq:seir_leq} given the test datasets in Table \ref{tab:leq_illustrative}.
As can be seen, the differences in each term of \eqref{eq:seir_leq} is less than $10^{-4}$, which might be negligible. 
This implies that the naïve approach of PINNs based on \eqref{eq:loss_M}, given partial measurements, approximates a certain solution to \eqref{eq:seir} with a different initial condition and parameter $\epsilon$, which is different from the ground truth. 
Regarding the initial conditions, one option for suppressing $L_\mathrm{init}(\theta_\mathrm{nn}; \cdot)$ is to adjust the constants $(\lambda_\mathrm{eq}, \lambda_\mathrm{data}, \lambda_\mathrm{init})$ such that $\lambda_\mathrm{eq} \ll  \lambda_\mathrm{init}$. However, this approach has two limitations. First, it reduces the effect of inductive bias $L_\mathrm{eq}(\theta_\mathrm{nn},\theta)$. This is not preferable owing to the limited availability of data. Second, hyperparameter tuning regarding the weights in the loss function of PINNs is typically equation- or data-dependent. 
Accordingly, in Section \ref{sec:proposed}, we propose an alternative approach to the inverse problem of PINNs given partial measurements.
\begin{table}
\caption{
Evaluation of $L_\mathrm{eq}(\theta_\mathrm{nn}, \theta)$ given the test dataset after the training based on \eqref{eq:loss_M}. All values are scaled by $10^{-6}$.
Each column corresponds to the evaluation of each term in \eqref{eq:seir_leq} in order.
\label{tab:leq_illustrative}
}
\begin{center}
\begin{tabular}{|c|c|c|c|c|}
\hline
\textbf{Measurements} & \textbf{$\dot{x}_1$}	&\textbf{$\dot{x}_2$} &\textbf{$\dot{x}_3$}   &\textbf{$\dot{x}_4$} \\
\hline
Partial
&$5.79$
&$4.24$
&$9.17$
&$28.94$\\
Full 
&$73.29$
&$3.82$
&$9.81$
&$58.22$ \\
\hline
\end{tabular}
\end{center}
\end{table}

\section{Algebraically observable PINNs}\label{sec:proposed} 
To overcome the difficulty in learning PINNs using partial measurements, we introduced the concept of algebraic observability \cite{meshkat}. 
In Section \ref{subsec:idea}, we explain the idea of the proposed method. 
In Section \ref{subsec:aoa}, algebraic observability analysis, which is a key technique in the proposed method, is introduced. Subsequently, the detailed description of the proposed method is provided in Section \ref{subsec:aopinn}.

\subsection{Concept of the proposed method}\label{subsec:idea}
As discussed in Section \ref{subsec:illustration}, partial measurements tend to hinder learning unmeasured state variables and unknown parameters. 
Considering this, we propose augmenting the data corresponding to the unmeasured state variables by utilising governing equations \eqref{eq:seir}. 
In the naive approach described in Section \ref{subsec:illustration}, \eqref{eq:seir} is introduced into the loss function \eqref{eq:loss_M} as an inductive bias. In addition, the proposed approach leverages the equations more effectively, particularly when generating augmented data. For this purpose, we introduce algebraic observability analysis. 

Intuitively, a state variable $x_i$ is algebraically observable if and only if a polynomial equation exists, such that
\begin{itemize}
    \item the set of variables comprises $x_i$, the measured state variables and the derivatives of the measured variables, and
    \item it lies in the set of polynomial equations obtained through differential and algebraic manipulations of \eqref{eq:model}.
\end{itemize}
For example, in the SEIR model with measurements of $I$ introduced in Section \ref{subsec:illustration}, the unmeasured state variable $E (= x_2)$ is algebraically observable. 
This is because the third equation in \eqref{eq:seir}, 
$\dot{I} = {\epsilon}{E}-{\gamma}{I}$, can be regarded as a polynomial equation of $E$, $I$, and $\dot{I}$, whose coefficients are functions of $(\beta, \gamma, \epsilon)$, considering the derivatives as independent of the variables. 
If an unmeasured state variable is algebraically observable, it can be represented using the measured state variable and its derivatives. 
For example, $E$ can be represented as follows:  
\begin{align}
\frac{{\dot{I}}+\gamma I}{\epsilon}, \label{eq:E_obs}
\end{align}
if $\epsilon \neq 0$.
Suppose that $\epsilon$ and $\gamma$ are known and the estimated value of the derivative of the measurement $\bar{I}(t_d)$, say $\dot{\bar{I}}(t_d)$, where $t_d \in \mathcal{T}$, is provided. 
A possible candidate for estimating $E(t_d)$ is
\begin{align}
\frac{\dot{\bar{I}}(t_d)+\gamma {\bar{I}}(t_d)}{\epsilon}
 \label{eq:E_hat}
\end{align}
according to Equation \eqref{eq:E_obs}. 
As such, in the proposed method, estimated values of unmeasured but algebraically observable variables are introduced as ``measurements'' during learning $\theta_\mathrm{nn}$ and $\theta$.
Specifically, the estimated values are incorporated into the term to evaluate deviations from the data. 
Owing to such augmented "measurements," the degree of freedom of the PINN is decreased, which is expected to result in regularising the PINNs. 

The following two points must be clarified to fully describe the proposed method.
The first describes the algebraic observability of the state variables. 
In particular, the derivation of polynomials required for estimating unmeasured variables and subsequently generating augmented data is critical. Therefore, we introduce an algebraic observability analysis framework \cite{meshkat}.  
The second problem is handling cases in which unknown parameters exist in such polynomials. In \eqref{eq:E_obs}, if either $\gamma$ or $\epsilon$ is unknown, then the estimation of $E$ is not straightforward anymore. 
Considering this, we combined Bayesian optimisation to deal with the unknown parameters. The details are provided in Section \ref{subsec:aopinn} along with the complete algorithm of the proposed method.

\subsection{Algebraic observability analysis}\label{subsec:aoa}
First, we review the algebraic terminology related to algebraic observability \cite{ritt,kolchin}. 
\begin{definition}
The differential ring $\mathcal{R}$ is a commutative ring with the derivatives $\partial:\mathcal{R} \rightarrow \mathcal{R}$. 
\end{definition}
In this study, we consider only a differential ring with a single derivative operator, in particular, $\mathrm{d}/\mathrm{d} t$.
\begin{definition}\label{definition:differentialideal}
The differential ideal $I$ of a differential ring $\mathcal{R}$ equipped with derivation operators $\partial$ is closed under the derivation.
\end{definition}
\begin{definition}
Let $\mathcal{R}$ be a differential ring equipped with the derivative $\mathcal{\partial}$. 
For a nonempty finite subset $p$ of $\mathcal{R}$, $p = \left\{p_1, \ldots, p_\lambda \right\}$, we define the differential ideal of $\mathcal{R}$ generated by $p$ as follows:
\begin{align}\label{eq:gen}
    \left\{ \sum_{i = 1}^\lambda
    \sum_{j = 0}^s
    c_{i, j}p_{i}^{(j)} \middle|  c_{i, j} \in \mathcal{R}, s \in \mathbb{N}
    \right\},\,
    p_i^{(j)} := \partial^j p_i\in  \mathcal{R}.
    \end{align}
We denote \eqref{eq:gen} as $[p]$ and call $p$ the generated set of $[p]$.
\end{definition}
Generally, the differential ideal generated by a finite subset $p \subset \mathcal{R}$ coincides with the minimal differential ideal containing $p$. Refer to \cite{kolchin} for further details. 

In this section, we consider the following polynomial equations:
\begin{align}
y = g(x; \theta_g), \label{eq:gmeasurement_model_} 
\end{align}
where $g$ denotes the function vector $\mathbb{R}^N \rightarrow \mathbb{R}^{M}$ and $g_i(x; \theta_g) \in \mathbb{Q}(\theta_g)[x]$ for all $i = 1, \ldots, M$. $\theta_g$ denotes a parameter vector. Here, $y = (y_1, \ldots, y_M)\in \mathbb{R}^{M}$ denotes the vector of measured variables depending on $t$. 
To simplify the notation, we denote the parameter vector of the state--space model comprising \eqref{eq:model} and \eqref{eq:gmeasurement_model} as $\theta \in \mathbb{R}^n$. Using this notation, the measurement equation \eqref{eq:gmeasurement_model_} can be represented as follows:
\begin{align}
y = g(x; \theta). \label{eq:gmeasurement_model} 
\end{align}
\begin{remark}\label{remark:pinn_general}
In Section \ref{subsec:pinn_inverse}, it is assumed that certain state variables are measured. Hence, the term that evaluates the deviation in the measured data is represented by \eqref{eq:ldataM}. 
As mentioned above, the measurement equation considered in the algebraic observability analysis is represented by \eqref{eq:gmeasurement_model}, which is more general. 
Using this notation, the measurement equation considered in Section \ref{subsec:pinn_inverse} can be represented as $g_i(x; \theta) = x_{\mathrm{idx}(i)}$, where $i \in \mathcal{M}$ and $M = |\mathcal{M}|$. 
Alternatively, the term that evaluates the deviation of the measured data in the loss function of PINNs \eqref{eq:loss_M} can be generalised as follows:
\begin{align}
\begin{split}
&L_{y}(\theta_\mathrm{nn}, \theta; \bar{x}_0, u, \bar{y}^\mathcal{T})\\
&= \lambda_{\mathrm{eq}}L_{\mathrm{eq}}(\theta_\mathrm{nn}, \theta; u)  
+ \lambda_{\mathrm{init}}L_{\mathrm{init}}(\theta_\mathrm{nn}; \bar{x}_0) 
+ \lambda_{\mathrm{data}}L_{\mathrm{data}}(\theta_\mathrm{nn}; \bar{y}^\mathcal{T}), 
\end{split}\label{eq:loss_y}
\end{align}
where the first and second terms are identical to those in \eqref{eq:loss_M}. The loss term is defined as follows:
\begin{align}
L_\mathrm{data}(\theta_\mathrm{nn}, \theta; \bar{y}^\mathcal{T}) 
&= \sum_{i = 1}^M
\left(\frac{1}{D}\sum_{d=1}^{D}{
\Big(
g_i\big({x}_{\mathrm{nn}}(t_d;\theta_\mathrm{nn}); \theta\big) - \bar{y}_i(t_d))
\Big)
}^2\right),\label{eq:ldata_general}
\end{align}
where $\bar{y}^\mathcal{T}$ denotes the measured data 
\begin{align}
\left\{
\begin{aligned}
\bar{y}(t_d)  
&= {
\left( 
\bar{y}_1(t_d), 
\cdots, 
\bar{y}_M(t_d)
\right)}^\top \\
&:= 
{\left( 
g_1\big( x(t_d; \theta_*, \bar{x}_0, u); \theta \big), 
\cdots, 
g_M\big( x(t_d; \theta_*, \bar{x}_0, u); \theta \big)
\right)}^\top 
\end{aligned}
\;\middle|\; t_d \in \mathcal{T} 
\right\}.
\label{eq:obs_y}
\end{align}
Note that \eqref{eq:ldata} can be considered a special case of \eqref{eq:ldata_general} by setting $g_i(x; \theta) = x_{\mathrm{idx}(i)}$ for all $i \in \mathcal{M}$ where $M = |\mathcal{M}|$.
\end{remark}

\begin{definition}[Algebraic observability, Definition 6.1 in \cite{meshkat}]\label{definition:ao}
Consider a state--space model comprising \eqref{eq:model} and \eqref{eq:gmeasurement_model}. 
The state variable $x_i$ is algebraically observable: for a given generic parameter vector $\theta$, there exists an open neighbourhood $X_i$ of the trajectory of $x_i(t)$ such that there is no other trajectory $\tilde{x}_i(t) \subseteq X_i$ that is compatible with the input--output trajectories $u(t), y(t)$.
\end{definition}

If $x_i$ is not algebraically observable, then it is defined as algebraically unobservable. The reason why this is known as "algebraically" observable is revealed in Proposition \ref{prop:prop_gb}. The algebraic observability defined in Definition \ref{definition:ao} is generically observable locally in Definition 6.1 of \cite{meshkat}.

In the following, we regard \eqref{eq:model}, \eqref{eq:gmeasurement_model} as elements in the differential ring $\mathbb{Q}[\theta]\{x,y,u\}$, of which the field is $\mathbb{Q}(\theta)$ and the indeterminants are $x, y, u$, where $\frac{\mathrm{d} \theta}{\mathrm{d} t} = 0$. The differential ideals generated by \eqref{eq:model} and \eqref{eq:gmeasurement_model} are considered as follows:
\begin{align}
\begin{split}
&[\dot{x}_1 - f_1(x,u;\theta), \ldots, \dot{x}_N - f_N(x,u;\theta), y_1 - g_1(x; \theta), \ldots, y_M - g_M(x; \theta)] \\
&\subset \mathbb{Q}(\theta)\left\{x,y,u\right\}. \label{eq:model_dideal}
\end{split} 
\end{align}

\begin{proposition}[proposition 6.2 in \cite{meshkat}]\label{prop:prop_h}
Consider a state--space model comprising \eqref{eq:model} and \eqref{eq:gmeasurement_model}. 
Let $I$ be the differential ideal \eqref{eq:model_dideal}. 
Let 
\begin{align*}
    h \in I \cap \mathbb{Q}(\theta)[x_i, y, \dot{y}, \ldots, u, \dot{u}, \ldots]
\end{align*}
be a polynomial, and write this as 
\begin{align}
h = \sum_{j=0}^k h_jx_i^j, \label{eq:sum_observable},
\end{align}
where the following three conditions hold:
\begin{itemize}
    \item $h_j \in \mathbb{Q}(\theta)[y, \dot{y}, \ldots, u, \dot{u}, \ldots]$ for $j = 0, \ldots, k$,
    \item $k \geq 1$,
    \item $h_k \notin I$.
\end{itemize}
Subsequently, $x_i$ is algebraically observable. 
If there is no polynomial 
\begin{align*}
h \in I \cap \mathbb{Q}(\theta)[x_i, y,\dot{y}, \ldots, u, \dot{u}, \ldots ]
\end{align*}
That satisfies these three conditions, then $x_i$ is algebraically unobservable.
\end{proposition}
According to Proposition \ref{prop:prop_h}, to determine whether $x_i$ is algebraically observable or not, it suffices to check whether there exists a polynomial {\eqref{eq:sum_observable}} in $I \cap \mathbb{Q}(\theta)[x_i, y,\dot{y},\ldots, u, \dot{u},\ldots]$ such that the three conditions hold.
In other words, to test whether $x_i$ is algebraically observable, we check whether the state variables, except for $x_i$ and their higher-order derivatives, can be eliminated from \eqref{eq:model}, \eqref{eq:gmeasurement_model}, and their derivatives. 

The following proposition suggests a strategy for algebraic observability analysis as well as for explicitly describing $h$ if it exists. 
\begin{proposition}[Proposition 6.3 in \cite{meshkat}]\label{prop:prop_gb}
We consider a state--space model comprising \eqref{eq:model} and \eqref{eq:gmeasurement_model}, where $M=1$. 
Let $J$ be the ideal
\begin{align}
\begin{split}
\langle 
&\dot{x} - f(x,u;\theta), \ldots, \frac{\mathrm{d}^{N-1} x}{\mathrm{d} t^{N-1}} - \frac{\mathrm{d}^{N-2}}{\mathrm{d} t^{N-2}}f(x,u;\theta), \\ 
&y - g(x; \theta), \ldots, y^{(N-1)} - \frac{\mathrm{d}^{N-1}}{\mathrm{d} t^{N-1}}g(x; \theta)\rangle \\
&\subset \mathbb{Q}(\theta)\left[x,y,u,\dot{x}, \dot{y}, \dot{u}, \ldots, \frac{\mathrm{d}^{N-2} x}{\mathrm{d} t^{N-2}}, \frac{\mathrm{d}^{N-2} y}{\mathrm{d} t^{N-2}}, \frac{\mathrm{d}^{N-2} u}{\mathrm{d} t^{N-2}},\frac{\mathrm{d}^{N-1} x}{\mathrm{d} t^{N-1}}, \frac{\mathrm{d}^{N-1} y}{\mathrm{d} t^{N-1}}\right].
\end{split}\label{eq:model_ideal}
\end{align}
Consider the elimination order $<_i$ of the 
\begin{align*}
\mathbb{Q}(\theta)\left[x,y,u,\dot{x}, \dot{y}, \dot{u}, \ldots, \frac{\mathrm{d}^{N-2} x}{\mathrm{d} t^{N-2}}, \frac{\mathrm{d}^{N-2} y}{\mathrm{d} t^{N-2}}, \frac{\mathrm{d}^{N-2} u}{\mathrm{d} t^{N-2}},\frac{\mathrm{d}^{N-1} x}{\mathrm{d} t^{N-1}}, \frac{\mathrm{d}^{N-1} y}{\mathrm{d} t^{N-1}}\right]
\end{align*}
using three blocks of variables 
\begin{align*}
\left\{x, \dot{x}, \ldots, \frac{\mathrm{d}^{N-1} x}{\mathrm{d} t^{N-1}}\right\} \backslash \left\{x_i\right\}
>
\left\{x_i\right\}
>
\left\{y, u, \dot{y}, \dot{u}, \ldots, \frac{\mathrm{d}^{N-2} y}{\mathrm{d} t^{N-2}}, \frac{\mathrm{d}^{N-2} u}{\mathrm{d} t^{N-2}},\frac{\mathrm{d}^{N-1} y}{\mathrm{d} t^{N-1}}\right\}.
\end{align*}
Then, a Gr\"{o}bner basis \cite{iva} for $J$ with respect to $<_i$ contains a polynomial $h$ represented as \eqref{eq:sum_observable}, which satisfies the three conditions listed in Proposition \ref{prop:prop_h} if it exists. Otherwise, no such polynomial $h$ exists; that is, $x_i$ is algebraically unobservable.
\end{proposition}
Proposition \ref{prop:prop_gb} shows that the algebraic observability analysis of the state--space models \eqref{eq:model} and \eqref{eq:gmeasurement_model}, where $M=1$ can be reduced to a Gr\"{o}bner basis computation.
In particular, it is reduced to eliminating the variables
\begin{align*}
\left\{x, \dot{x}, \ldots, \frac{\mathrm{d}^{N-1} x}{\mathrm{d} t^{N-1}}\right\} \backslash \left\{x_i\right\}
\end{align*}
from a finite set of polynomial equations: 
\begin{align*}
\begin{split}
&\dot{x} - f(x,u;\theta), \ldots, \frac{\mathrm{d}^{N-1} x}{\mathrm{d} t^{N-1}} - \frac{\mathrm{d}^{N-2}}{\mathrm{d} t^{N-2}}f(x,u;\theta), \\
&y - g(x; \theta), \ldots, y^{(N-1)} - \frac{\mathrm{d}^{N-1}}{\mathrm{d} t^{N-1}}g(x; \theta)
\end{split}
\end{align*}
by using the Gr\"{o}bner basis of $J$ with respect to $<_i$. 
Note that the order of the derivatives of $y, u$ in Eq. \eqref{eq:model_ideal} is bounded. Denoting the highest-order derivatives of $y, u$ as $p, q$, it holds that $p \leq N-1, q \leq N-2$. 
Proposition \ref{prop:prop_gb} can be extended to the case in which $M > 1$. See \cite{meshkat} for further details. Although it is possible to consider the case with $M > 1$, theoretically, we focus on the case with $M = 1$ as follows: See \ref{sec:conclusion} for further discussion.  

In general, the elimination process based on the Gr\"{o}bner basis can be computed using computer algebra software, for example, Singular \cite{singular}.
See, for example, \cite{iva} and the Appendices of \cite{robot} for an introductory explanation of elimination based on the Gr\"{o}bner basis computation. 
Owing to the algebraic computations, we can explicitly describe how to estimate unmeasured but algebraically observable variables. 
In particular, if the Gr\"{o}bner basis computed in Proposition \ref{prop:prop_gb} contains $h$ satisfying the three conditions, then a solution of $h=0$ with respect to $x_i$ exists. Subsequently, this can be used as an estimation function of $x_i$. 

The set of indices of variables that are unmeasured and algebraically observable is denoted as $\mathcal{A} \subseteq \mathcal{N}\backslash \mathcal{M}$.  
For $i \in \mathcal{A}$, let $h$ take the form \eqref{eq:sum_observable} 
\begin{align}
H_i(x_i, y, \dot{y}, \ldots, u, \dot{u}, \ldots; \theta). \label{eq:Hi}
\end{align} 
In the following, we assume that the higher-order derivatives of $u$ at $t \in \mathcal{T}$ that appear in $H_i$ are known. 
This is because the solution of $H_i = 0$ with respect to $x_i$ can be represented as a function of $y, u$, and their higher derivatives. 
The treatment of higher derivatives of $y$ is presented in Section \ref{subsec:aopinn}.

\begin{example}\label{example:seir_ao_ex}
In this example, we demonstrate an algebraic observability analysis of the SEIR model by using the measurement equation $y = I$. See Section \ref{sec:singular_ex} for the singular commands corresponding to this example.
Here, we introduce $x = {(S, E, I)}^\top \in \mathbb{R}^3$. Thus, \eqref{eq:seir}, without the last equation, is expressed as follows:
\begin{align}
\begin{split}
\dot{x}_1 = -{\beta}{x_1}x_3,\quad 
\dot{x}_2 = {\beta}{x_1}{x_3}-{\epsilon}{x_2},\quad
\dot{x}_3 = {\epsilon}{x_2}-{\gamma }{x_3}.
\end{split}\label{eq:seir_x}
\end{align}
$R$ is omitted from \eqref{eq:seir_x}, because it holds that $R = 1 -(S+ E+I)$. 
In line with Section \ref{subsec:illustration}, we assume that measurements of $I(=x_3)$ are provided, which leads to the following measurement equation:
\begin{align}
y = I = x_3.\label{eq:seir_measure}
\end{align}
Therefore, $\mathcal{M} = \left\{ 3\right\}, \mathcal{N} = \left\{1, 2,3\right\}$. Accordingly, the set of indices for unmeasured state variables is represented by $\mathcal{N}\backslash \mathcal{M} = \left\{1, 2 \right\}$. 

Next, we verify the algebraic observability of $x_i$, where $i \in \mathcal{N}\backslash \mathcal{M}$. 
For $x_2$, we consider the ideal 
\begin{align*}
J_1 \subset \mathbb{Q}(\beta, \epsilon, \gamma )\left[x_1, x_2, x_3, y,\dot{x}_1, \dot{x}_2, \dot{x}_3, \dot{y}, \ddot{x}_1, \ddot{x}_2, \ddot{x}_3, \ddot{y}\right]
\end{align*}
as follows:
\begin{align*}
\begin{split}
J_1 = \langle &
\dot{x}_1 + {\beta}{x_1}x_3, 
\dot{x}_2 -{\beta}{x_1}{x_3}+{\epsilon}{x_2}, \dot{x}_3 - {\epsilon}{x_2}+{\gamma }{x_3},  
\ddot{x}_1 + {\beta}\left(\dot{x}_1x_3 + x_1\dot{x}_3\right), \\
&\ddot{x}_2 -{\beta}\left(\dot{x}_1x_3 + x_1\dot{x}_3\right)+{\epsilon}{\dot{x}_2}, \ddot{x}_3 - {\epsilon}{\dot{x}_2}+{\gamma }{\dot{x}_3},
y - x_3, \dot{y} - \dot{x}_3, \ddot{y} - \ddot{x}_3 
\rangle.
\end{split}
\end{align*}
Using Proposition \ref{prop:prop_gb}, we consider the intersection $J_1 \cap \mathbb{Q}(\beta, \epsilon, \gamma )\left[x_2, y, \dot{y}, \ddot{y}\right]$ and check for the existence of a polynomial $h = \sum_{j=0}^k h_jx_2^j$
where we satisfy the three conditions in Proposition \ref{prop:prop_h}. 
We consider lexicographic ordering $<_2$, which is one of the elimination orders for 
\begin{align*}
    \mathbb{Q}(\beta, \epsilon, \gamma )\left[x_1, x_2, x_3, y, \dot{x}_1, \dot{x}_2, \dot{x}_3, \dot{y}, \ddot{x}_1, \ddot{x}_2, \ddot{x}_3, \ddot{y}\right]
\end{align*}
as follows:
\begin{align*}
\left\{x_1, x_3, \dot{x}_1, \dot{x}_2, \dot{x}_3, \ddot{x}_1, \ddot{x}_2, \ddot{x}_3 \right\}
>
\left\{x_2\right\}
>
\left\{y, \dot{y}, \ddot{y}\right\}.
\end{align*}
An example of the series of singular commands used to compute the Gr\"{o}bner basis is provided in Appendix \ref{sec:singular_ex}. 
The Gr\"{o}bner basis of $J_1$ with respect to $<$ contains the following elements.
\begin{align*}
\epsilon x_2-\gamma y-\dot{y} \in J_1 \cap \mathbb{Q}(\beta, \epsilon, \gamma )\left[x_2, y, \dot{y}, \ddot{y}\right]. 
\end{align*}
Denoting this polynomial as $h$, it holds that $h_0 = -(\gamma y+\dot{y}), h_1 = \epsilon$. 
This satisfies the following three conditions:
\begin{itemize}
    \item $h_0, h_1 \in \mathbb{Q}(\beta, \epsilon, \gamma )[y,\dot{y}]$,
    \item $k = 1 \geq 1$,
    \item $h_1 \notin J_1$.
\end{itemize}
Thus, $x_2$ is algebraically observable. 

Similarly, to check the algebraic observability of $x_1$, we consider the intersection $J_1 \cap \mathbb{Q}(\beta, \epsilon, \gamma )\left[x_1, y, \dot{y}, \ddot{y}\right]$ and check for the existence of a polynomial $\sum_{j=0}^k h_jx_1^j$
satisfying these three conditions. 
The corresponding singular commands can only be obtained by changing the ring specifications. Specifically, the first two lines in Listing \ref{listing:ex1x2} are replaced by the commands in Listing \ref{listing:ex1x1}. The output is shown in Listing \ref{listing:ex1x1out} and includes 
\begin{align*}
(\beta \epsilon)x_1y+(-\epsilon \gamma )y+(-\epsilon-\gamma )\dot{y}-\ddot{y} \in J_1 \cap \mathbb{Q}(\beta, \epsilon, \gamma )\left[x_1, y, \dot{y}, \ddot{y}\right].
\end{align*}
As before, $x_1$ is observable algebraically. 
Finally, we obtain the following:
\begin{align}
\mathcal{A} &= \left\{1, 2\right\} = \mathcal{N} \backslash \mathcal{M}, \label{eq:seir_A}\\
H_1(x_1, y,\dot{y}, \ddot{y}; \theta) &= (\beta \epsilon)x_1y+(-\epsilon \gamma )y+(-\epsilon-\gamma )\dot{y}-\ddot{y}, \label{eq:seir_H1} \\
H_2(x_2, y,\dot{y}, \ddot{y}; \theta) &= \epsilon x_2-\gamma y-\dot{y}.   \label{eq:seir_H2} 
\end{align}
In this example, the unmeasured and the algebraically observable variables coincide. As both $x_1$ and $x_2$ are algebraically observable, $R = 1 -(S+E+I) = 1- (x_1 + x_2+ x_3)$ can also be represented as a function of $u, y$ and their higher-order derivatives.
\end{example}
\subsection{Proposed method: Algebraically observable PINN}\label{subsec:aopinn}
We assume that the measurement time points $\mathcal{T}$ are divided into three sets: the sets of time points used for training, validation, and testing. They are denoted as $\mathcal{T}_\mathrm{train}, \mathcal{T}_\mathrm{val}$ and $\mathcal{T}_\mathrm{test}$ where $\mathcal{T} = \mathcal{T}_\mathrm{train} \cup \mathcal{T}_\mathrm{val} \cup \mathcal{T}_\mathrm{test}$. Using these notations, the partial measurements for the training, validation, and test datasets are denoted by $y^{{\mathcal{T}}_\mathrm{train}}, y^{{\mathcal{T}}_\mathrm{val}}, y^{{\mathcal{T}}_\mathrm{test}}$, respectively. 
In addition, we make the following implicit assumptions:
\begin{assumption}
Suppose we have government equations \eqref{eq:model}, measurement equations \eqref{eq:gmeasurement_model}, and partial measurements $y^{\mathcal{T}}$. For all $t_d \in \mathcal{T}$, we assume that the values of $u^{(1)}(t_d), \ldots, u^{(q)}(t_d)$ can be estimated based on $u$: We also assume that the values of $y^{(1)}(t_d), \ldots, y^{(p)}(t_d)$ can be estimated by using $y^\mathcal{T}$:
\end{assumption}
As the input function $u$ is known and sufficiently smooth, the first assumption is not strong; in fact, exact values may even be available.
Regarding the latter, several methods for estimating higher-order derivatives based on the measurements of a function are known, such as AAA \cite{AAAalg} and Sobolev learning \cite{sobolev}. 
In the following, we denote the estimated values of $u^{(1)}(t_d), \ldots, u^{(q)}(t_d)$ and $y^{(1)}(t_d), \ldots, y^{(p)}(t_d)$ as $\tilde{u}^{(1)}(t_d), \ldots, \tilde{u}^{(q)}(t_d)$ and $\tilde{y}^{(1)}(t_d), \ldots, y^{(p)}(t_d)$, respectively,

Next, we explain the details of the proposed method. See Algorithm \ref{algorithm:proposedalg} for details of the proposed method. 
As explained in Section \ref{subsec:partial}, we consider the inverse problem of PINNs, whose governing equations are \eqref{eq:model}, given the partial measurements. Based on Remark \ref{remark:pinn_general}, we consider partial measurements \eqref{eq:E_obs} and measurement equations \eqref{eq:gmeasurement_model}. 

As the first step in the proposed method, we perform the algebraic observability analysis explained in Section \ref{subsec:aoa}. 
Consequently, we obtain a set of indices of unmeasured but algebraically observable variables $\mathcal{A}$ and a set of polynomials $\left\{ H_i\mid i \in \mathcal{A}\right\}$. 

Then, the following procedure is repeated for $s = 1, 2, \ldots, S$ where $S$ is an integer representing the number of samples of the parameter vectors. The parameter vector $\theta^{(s)} \sim P_\theta$ is sampled. Here, $P_\theta$ denotes the probability distribution of the parameter vectors. 
We estimate the unmeasured but algebraically observable variables $x_i (i \in \mathcal{A})$ given $\theta = \theta^{(s)}$.
Specifically, let us denote the solution of $H_i = 0$ with respect to $x_i$ as 
\begin{align}
\tilde{H}_i
\left(
y, y^{(1)}, \ldots, y^{(p)}, u, u^{(1)}, \ldots, u^{(q)}; \theta
\right).\label{eq:tHi}
\end{align}
Given $\theta = \theta^{(s)}$, the values of $x_i(t_d)$, where $i \in \mathcal{A}$ and $t_d \in \mathcal{T}$, are estimated as follows:
\begin{align}
\tilde{H}_i\left(
{y}(t_d), \tilde{y}^{(1)}(t_d), \ldots, \tilde{y}^{(p)}(t_d), {u}(t_d), \tilde{u}^{(1)}(t_d), \ldots, \tilde{y}^{(q)}(t_d); \theta^{(s)}
\right). \label{eq:estfun}
\end{align}
We denote this as $\hat{x}_i(t_d; \theta^{(s)})$ explicitly shows the dependency on $\theta^{(s)}$. 
Consequently, we obtain the following augmented data, denoted by $\bar{x}_\mathcal{A}^\mathcal{T}$:
\begin{align}
\left\{
\begin{aligned}
\bar{x}_\mathcal{A}(t_d)  
&= {
\left( 
\bar{x}_{\mathrm{idx}(1)}(t_d), 
\cdots, 
\bar{x}_{\mathrm{idx}(|\mathcal{A}|)}(t_d)
\right)}^\top \\
&:= 
{\left( 
\hat{x}_{\mathrm{idx}(1)}(t_d; \theta^{(s)}), 
\cdots, 
\hat{x}_{\mathrm{idx}(|\mathcal{A}|)}(t_d; \theta^{(s)})
\right)}^\top 
\end{aligned}
\;\middle|\; t_d \in \mathcal{T} 
\right\}.
\label{eq:augmented}
\end{align}
By introducing augmented data, we define the loss function as follows:
\begin{align}
\begin{split}
&L_{y, \mathcal{A}}\left(\theta_\mathrm{nn}; \theta^{(s)}, \bar{x}_0, u, \bar{y}^{\mathcal{T}_\mathrm{train}}, \bar{x}_\mathcal{A}^{\mathcal{T}_\mathrm{train}}\right)\\
&= L_{y}\left(\theta_\mathrm{nn}, \theta^{(s)}; \bar{x}_0, u, \bar{y}^{\mathcal{T}_\mathrm{train}}\right) 
+ \lambda_{\mathrm{data}}L_{\mathrm{data}}\left(\theta_\mathrm{nn}; \bar{x}_\mathcal{A}^{\mathcal{T}_\mathrm{train}}\right), 
\end{split}\label{eq:loss_aug}
\end{align}
where $L_{\mathrm{data}}(\theta_\mathrm{nn}; \bar{x}_\mathcal{A}^{\mathcal{T}_\mathrm{train}})$ is defined by replacing $\mathcal{M}$ with $\mathcal{A}$ and $\mathcal{T}$ with $\mathcal{T}_\mathrm{train}$ in the right side of \eqref{eq:ldataM}, that is, 
\begin{align}
L_{\mathrm{data}}\left(\theta_\mathrm{nn}; \bar{x}_\mathcal{A}^{\mathcal{T}_\mathrm{train}}\right)
:= \sum_{i \in \mathcal{A}}
\left(
\frac{1}{|\mathcal{T}_\mathrm{train}|}\sum_{d=1}^{|\mathcal{T}_\mathrm{train}|}{({x}_{\mathrm{nn},i}(t_d;\theta_\mathrm{nn}) - \bar{x}_i(t_d))}^2
\right).\label{eq:ldataA}
\end{align}
Note that $L_{y}(\theta_\mathrm{nn}, \theta^{(s)}; \bar{x}_0, u, \bar{y}^{\mathcal{T}_\mathrm{train}})$ is the one defined in \eqref{eq:loss_y} where $\theta$ is fixed to $\theta^{(s)}$ and $\mathcal{T}$ is replaced with $\mathcal{T}_\mathrm{train}$.
Subsequently, $\theta_\mathrm{nn} \in \mathbb{R}^{n'}$ is trained as follows:
\begin{align}
\min_{\theta_\mathrm{nn}} L_{y, \mathcal{A}}\left(\theta_\mathrm{nn}; \theta, \bar{x}_0, u, \bar{y}^{\mathcal{T}_\mathrm{train}}, \bar{x}_\mathcal{A}^{\mathcal{T}_\mathrm{train}}\right). \label{eq:min_loss_aug}
\end{align}
Based on \eqref{eq:min_loss_aug}, we obtain 
\begin{align}
\theta_{\mathrm{nn}}^{(s)} := \argmin_{\theta_{\mathrm{nn}}} L_{y, \mathcal{A}}\left(\theta_\mathrm{nn}; \theta, \bar{x}_0, u, \bar{y}^{\mathcal{T}_\mathrm{train}}, \bar{x}_\mathcal{A}^{\mathcal{T}_\mathrm{train}}\right). \label{eq:agrmin_theta_nn}
\end{align}
Using the validation data $\bar{y}^{\mathcal{T}_\mathrm{val}}$, we obtain the loss function \eqref{eq:loss_aug}, given $\theta_{\mathrm{nn}}^{(s)}$. 
Specifically, we evaluate the following:
\begin{align}
E_\mathrm{val}^{(s)}
:=
L_{y, \mathcal{A}}\left(
\theta_{\mathrm{nn}}^{(s)}; \theta^{(s)}, \bar{x}_0, u, \bar{y}^{\mathcal{T}_\mathrm{val}}, \bar{x}_\mathcal{A}^{\mathcal{T}_\mathrm{val}}
\right). \label{eq:obj_val}
\end{align}
Upon completion of the final iteration, $s = S$, we obtain 
\begin{align}
s_* &:= \argmin_s E_\mathrm{val}^{(s)}.
\end{align}
Accordingly, we obtain $(\hat{\theta}_\mathrm{nn}, \hat{\theta})$ as 
\begin{align}
\hat{\theta} := \theta^{(s_*)}, \quad 
\hat{\theta}_\mathrm{nn} := \theta_\mathrm{nn}^{(s_*)}.
\end{align}
Hence, the state and parameter estimations of \eqref{eq:model} and \eqref{eq:gmeasurement_model} are completed, and the learned state vector and estimated parameter vector correspond to $x_\mathrm{nn}(t; \hat{\theta}_\mathrm{nn})$ and $\hat{\theta}$, respectively.   
A neural network trained with the loss function \eqref{eq:loss_aug} is referred to as an algebraically observable PINN.
\begin{algorithm}[H]
    \caption{State and parameter estimation of \eqref{eq:model} and \eqref{eq:gmeasurement_model} based on algebraically observable PINN.}
    \label{algorithm:alg1}
    \begin{algorithmic}[1] 
    \REQUIRE \eqref{eq:model}, \eqref{eq:gmeasurement_model}, known input function $u$, initial conditions $x(0) = \bar{x}_0$, (partial) measurements as training and validation data set $y^{\mathcal{T}_\mathrm{train}}, y^{\mathcal{T}_\mathrm{val}}$, the neural network $x_\mathrm{nn}(t; \theta_\mathrm{nn})$, the probability distribution $P_\theta$.
    \ENSURE Estimated state variables $x_\mathrm{nn}(t; \hat{\theta}_\mathrm{nn})$, estimated parameters $\hat{\theta}$. 
    \STATE {Perform algebraic observability analysis for \eqref{eq:model}, \eqref{eq:gmeasurement_model}, which outputs $\mathcal{A}$ and $\left\{H_i \mid i \in \mathcal{A}\right\}$. }
    \STATE {Based on $y^{\mathcal{T}_\mathrm{train}}$ and $ y^{\mathcal{T}_\mathrm{val}}$, estimate the values of $\left\{ y^{(k)}(t_d) \mid t_d \in \mathcal{T}_\mathrm{train} \cup \mathcal{T}_\mathrm{val} \right\}$ for $k = 1, \ldots, p$. The estimated values are represented as variables with tildes.}
    \STATE {Based on the input function $u$, estimate the values of $\left\{ u^{(k)}(t_d) \mid t_d \in \mathcal{T}_\mathrm{train} \cup \mathcal{T}_\mathrm{val}\right\}$ for $k = 1, \ldots, q$. The estimated values are represented as variables with tildes.}
    \FOR{$s = 1, \ldots, S$}
    \STATE Sample $\theta^{(s)} \sim P_\theta$.
    \STATE Generate the augmented data $x_\mathcal{A}^{\mathcal{T}_\mathrm{train}}$ and $ x_\mathcal{A}^{\mathcal{T}_\mathrm{val}}$ based on \eqref{eq:estfun} and \eqref{eq:augmented}.
    \STATE $\theta_{\mathrm{nn}}^{(s)} \leftarrow \argmin_{\theta_\mathrm{nn}} L_{y, \mathcal{A}}\left(\theta_\mathrm{train}; \theta, \bar{x}_0, u, \bar{y}^{\mathcal{T}_\mathrm{train}}, \bar{x}_\mathcal{A}^{\mathcal{T}_\mathrm{train}}\right)$.
    \STATE $E_\mathrm{val}^{(s)} \leftarrow L_{y, \mathcal{A}}\left(
    \theta_{\mathrm{nn}}^{(s)}; \theta^{(s)}, \bar{x}_0, u, \bar{y}^{\mathcal{T}_\mathrm{val}}, \bar{x}_\mathcal{A}^{\mathcal{T}_\mathrm{val}}
    \right).$
    \ENDFOR
    \STATE $s_* \leftarrow \argmin_s E_\mathrm{val}^{(s)}$.
    \STATE $(\hat{\theta}_\mathrm{nn}, \hat{\theta}) \leftarrow ({\theta_\mathrm{nn}}^{(s_*)}, \theta^{(s_*)})$.
    \RETURN $x_\mathrm{nn}(t, \hat{\theta}_\mathrm{nn}), \hat{\theta}$
    \end{algorithmic}\label{algorithm:proposedalg}
\end{algorithm}

\section{Numerical Experiments}\label{sec:exp}
As a proof of concept, we considered three target scenarios. In each scenario, the numerical solution of an epidemiological model was used as the ground truth, and partial measurements were obtained based on it. The target scenarios are described below. Specifically, for each target scenario, we provide details of the epidemiological model, which serves as the governing equations, measurement equations, initial conditions, input functions, and epidemiological parameters to be estimated, along with their values. 
In addition, the results of algebraic observability analysis are presented.
Subsequently, the details of the dataset preparation and the training and evaluation of the PINNs are provided. 
\subsection{Target scenario 1: SEIR model}\label{subsec:target_seir}
For Target scenario 1, we follow the situation in Section \ref{subsec:illustration}. The SEIR model \eqref{eq:seir} is used as the governing equation and \eqref{eq:seir_measure} as the measurement equation. 
Given the initial conditions $(S(0), E(0), I(0), R(0)) = (0.99, 0.0, 0.01, 0.0)$ and fixed parameters $(\beta, \gamma) = (0.26, 0.1)$, the values of $\epsilon$ are estimated, where the true value $\epsilon_*$ is set to 0.2. Note that there is no input function in this scenario.

An algebraic observability analysis of this scenario is presented in Example \ref{example:seir_ao_ex}. Specifically, for $x = {(x_1, x_2, x_3)}^\top = {(S, E, I)}^\top \in \mathbb{R}^3$, we obtain \eqref{eq:seir_A}, \eqref{eq:seir_H1}, and \eqref{eq:seir_H2}. 
The solution to \eqref{eq:seir_H1} with respect to $x_1$ is 
\begin{align}
\tilde{H}_1(y, \dot{y}, \ddot{y}; \beta, \gamma, \epsilon) := 
\frac{\epsilon \gamma y + (\epsilon + \gamma)\dot{y}+ \ddot{y}}{\beta \epsilon y},
\end{align}
and \eqref{eq:seir_H2} with respect to $x_2$ are
\begin{align}
\tilde{H}_2(y, \dot{y}, \ddot{y}; \beta, \gamma, \epsilon) := 
\frac{\gamma y + \dot{y}}{\epsilon}.
\end{align}

\subsection{Target scenario 2: SICRD model}\label{subsec:target_sicrd}
For Target scenario 2, we considered the SICRD model \cite{sicrd} as a government equation: 
\begin{align}
\begin{split}
\dot{S} &= -\beta SI-qS+pC,\\
\dot{I} &= \beta SI - (r+\mu)I,\\
\dot{C} &= qS - pC, \\
\dot{R} &= rI,\\
\dot{D} &= \mu I, 
\end{split}\label{eq:sicrd}
\end{align}
where the values of $(p, q, r, \mu)$ are fixed to $(0.01, 0.01, 0.05, 0.05)$.
The initial conditions are set to $(S(0), I(0), C(0), R(0), D(0)) = (0.99, 0.01, 0.0, 0.0, 0.0)$. The parameter to be estimated is $\beta$, where the true value $\beta_*$ is set to 0.26. 
We assume the measurement of $I$. 
To check the algebraic observability of $S, C$, we set $x = {(x_1, x_2, x_3)}^\top = {(S, I, C)}^\top$ and consider the following measurement equation: 
\begin{align}
y_1 = I = x_2.\label{eq:sicrd_measure_ao}    
\end{align}
It holds that $\mathcal{N} = \left\{1, 2, 3, 4\right\}$ and $\mathcal{M} = \left\{2\right\}$.
The reason $D$ is omitted from $x$ is essentially the same as in Example \ref{example:seir_ao_ex}; it holds that $D = 1 - (S + I + C+ R)$.
As $R$ does not appear in \eqref{eq:sicrd}, we omit it if $x$; $R$ is obviously unobservable algebraically.
The results of the algebraic observability analysis are as follows:
\begin{align}
H_1(x_1, y_1, \dot{y}_1; \beta) =& 10\beta x_1y_1-10\dot{y}-y_1, \label{eq:sicrd_H1}\\
\begin{split}
H_3(x_3, y_1, \dot{y}_1, \ddot{y}_1; \beta, p, q) =& (10\beta p )x_3 y_1^2-10\ddot{y}_1y_1+10\dot{y}_1^2-10\beta\dot{y}_1y_1^2\\
&-10q\dot{y}_1y_1-\beta y_1^3-qy_1^2, 
\end{split}\label{eq:sicrd_H3}\\
\mathcal{A} =& \left\{
1, 3
\right\}. \label{eq:sicrd_A}
\end{align}
To reduce the cost of computing the Gr\"{o}bner basis, the values or $(r, \mu)$ are substituted into \eqref{eq:sicrd} before the analysis.
The solution to \eqref{eq:sicrd_H1} with respect to $x_1$ is 
\begin{align}
\tilde{H}_1(x_1, y_1, \dot{y}_1; \beta) = \frac{10\dot{y}_1+y_1}{10\beta y_1}
\end{align}
and the solution of \eqref{eq:sicrd_H3} with respect to $x_3$ is 
\begin{align*}
\tilde{H}_3(y_1, \dot{y}_1, \ddot{y}_1; \beta, p, q) = \frac{10\ddot{y}_1y_1-10\dot{y}_1^2+10\beta\dot{y}_1y_1^2+10q\dot{y}_1y_1+\beta y_1^3+qy_1^2}{10\beta p y_1^2}.
\end{align*}

\subsection{Target scenario 3: SAIRD model}\label{subsec:target_saird}
For Target scenario 3, we consider the SAIRD model \eqref{eq:saird} as the government equations, as defined in Example \ref{example:saird}. 
Notably, \eqref{eq:saird} has an input function $u = \exp(-kt)$, which differs from the other scenarios. 
We assume that the values of $(\xi, \gamma, \delta, k)$ are known and that $(\xi, \gamma, \delta, k) = (0.1, 0.05, 0.05, 0.01)$. The initial conditions are set as follows: 
\begin{align*}
    (S(0), A(0), I(0), R(0), D(0)) = (0.985, 0.005, 0.01, 0.0, 0.0).
\end{align*}
The parameter to be estimated are $(\beta, \kappa)$, where the true value $(\beta_*, \kappa_*)$ is set as $(0.26, 0.1)$. 
We assume that the measurements of $I, R$ are provided.
To verify the algebraic observability of $S, A$, we consider $x = {(x_1, x_2, x_3, x_4)}^\top = {(S, A, I,  R)}^\top$.
It holds that $\mathcal{N} = \left\{1, 2, 3, 4\right\}$ and $\mathcal{M} = \left\{3, 4\right\}$.
Then, the SAIRD model \eqref{eq:saird} without the last two equations, given the measurement equation
\begin{align}
y_1 = I = x_3\label{eq:saird_measure_ao}    
\end{align}
is analysed, as in Target scenario 2.
The results are as follows:
\begin{align}
H_1(x_1, y_1, \dot{y}_1, \ddot{y}_1, u; \beta, \xi) &= 100\xi x_1\dot{y}_1u+10(\beta +\xi)x_1y_1u-100\ddot{y}_1-20\dot{y}_1-y_1, \label{eq:saird_H1}\\
H_2(x_2, y_1, \dot{y}_1) &= x_2-10\dot{y}_1-y_1, \label{eq:saird_H2}\\
\mathcal{A} &= \left\{
1, 2
\right\} = \mathcal{N}\backslash \mathcal{M}. 
\label{eq:saird_A}
\end{align}
To reduce the computational cost, the values of $(\gamma, \delta)$ are substituted into \eqref{eq:saird} before computing the Gr\"{o}bner basis.
The solution to \eqref{eq:saird_H1} with respect to $x_1$ is 
\begin{align*}
\tilde{H}_1(y_1, \dot{y}_1, \ddot{y}_1, u; \beta, \xi) = 
\frac{100\ddot{y}_1+20\dot{y}_1+y_1}{100\xi\dot{y}_1u+10(\beta +\xi )y_1u},
\end{align*}
and the solution to \eqref{eq:saird_H2} with respect to $x_2$ is 
\begin{align*}
\tilde{H}_2(y_1, \dot{y}_1) = 10\dot{y}_1 + y_1.
\end{align*}
\subsection{Data preparation}\label{subsec:data}
This section describes the numerical simulations and datasets used in this study.
The epidemiological models were numerically solved over the period $[0, 200]$ given the initial conditions defined for each scenario. 
The Dormand--Prince method with a time-step size $\Delta t = 0.2$ was used as the numerical solver. 
For the training and test datasets, the measured time points were selected randomly from the evaluation points of the numerical solutions. The validation dataset was randomly sampled from a uniform distribution over $[0, 200]$. 
Regarding the size of each dataset, 
$|\mathcal{T}_\mathrm{train}| = |\mathcal{T}_\mathrm{val}| = 50, |\mathcal{T}_\mathrm{test}| = 100$. 
To evaluate the robustness of the methods, noise sampled from a uniform distribution over $[0, \sigma]$ was added to each state variable at each measurement time point for Target scenario 1. In particular, we consider two noise settings: $\sigma = 0.05$ and $\sigma = 0.1$.

As previously mentioned, the values of the higher derivatives of the measured variables at the measured time points are required for the proposed framework.
The primary focus of this study is on the effect of the augmented data generated based on algebraic observability analysis. 
Therefore, we did not consider the estimation accuracy of the higher-order derivatives across all scenarios. 
For this purpose, we analytically derived the higher-order derivatives of the right side of the epidemiological models and measurement equations, given a true parameter vector, and substituted the numerical solutions into them. Notably, the numerical solutions for target Scenario 1 included noise. 
This approach yields estimates of the higher-order derivatives of $y$, where the numerical error is the sole source of the estimation error.

\subsection{Experimental settings}\label{subsection:ex_setting}
In the numerical experiments for target scenario 1, we compared three methods: the proposed method explained in Section \ref{sec:proposed}, a reference method explained later, and the baseline explained in Section \ref{subsec:partial}. 
The proposed method is described in Algorithm \ref{algorithm:alg1}. 
The reference method is based on Algorithm \ref{algorithm:alg1}, but replaces the loss function $ L_{y, \mathcal{A}}\left(\theta_\mathrm{train}; \theta, \bar{x}_0, u, \bar{y}^{\mathcal{T}_\mathrm{train}}, \bar{x}_\mathcal{A}^{\mathcal{T}_\mathrm{train}}\right)$ 
with the one that corresponds to \eqref{eq:loss_y}, that is, $ L_{y}\left(\theta_\mathrm{nn}, \theta^{(s)}; \bar{x}_0, u, \bar{y}^{\mathcal{T}_\mathrm{train}}\right)$. 
Hence, the loss function used in the reference method does not incorporate the augmented data. The baseline method also does not incorporate augmented data. 
Although the loss functions used in the reference and baseline methods are the same, the learning methods $\theta_\mathrm{nn}, \theta$ are different: the baseline method learns $\theta_\mathrm{nn}$ and $\theta$ simultaneously, whereas the proposed and reference methods learn $\theta_\mathrm{nn}$ and $\theta$ separately by sampling the parameter vectors. 

In numerical experiments for Target Scenarios 2 and 3, we verified the effectiveness of the proposed method for more advanced problems. 
Specifically, for Target Scenario 2, the validity of the proposed method in the presence of unmeasured and algebraically unobservable variables was investigated and compared with the baseline method. 
In addition, we compared the results obtained using the proposed method with and without the initial condition of $R(0)$. 
The loss function \eqref{eq:loss_aug} for the former includes a deviation of $R(0) = x_4(0)$ in $L_\mathrm{init}(\theta_\mathrm{nn}; \bar{x}_0)$, whereas, the latter does not. 
Regarding Target Scenario 3, numerical experiments were conducted to assess the performance of the proposed method, given an input function. In addition, the effectiveness of the proposed method was investigated under multiple unknown parameters.

In the proposed and reference methods, we introduce the Gaussian process-based Bayesian optimisation (GP-BO) \cite{gpbo,hporev} for the sampling parameters $\theta^{s} (s = 1, \ldots, S)$. 
GP-BO is often applied to hyperparameter optimisation problems \cite{hpo} and is known to be effective in achieving good performance with a limited number of evaluations compared to grid search. 
In our context, the unknown parameter vector $\theta$, which is used to generate augmented data, is regarded as a hyperparameter. 
As the objective function of GP-BO, we used $E_\mathrm{val}^{(s)}$ defined in the eighth line of Algorithm \ref{algorithm:alg1}. 
In the experiments, the GP-BO with Expected Improvement \cite{EI,hpo} as the acquisition function was implemented. 
The number of iterations for GP-BO was set to $S = 30$ for Target Scenarios 1 and 2, and $S = 50$ for Target scenario 3. 
The search space of the parameter vector is set to ${[0.0, 0.5]}^n$, where $n$ denotes the number of parameters to be estimated.

Throughout the experiments, the same neural network architecture was used for all methods for comparison. Specifically, we selected a fully neural network with three hidden layers. Each hidden layer had 50 units that preceded the hyperbolic tangent activation function. The coefficients of the loss functions were set to 1s for all methods across scenarios.
The loss function was minimised with respect to $(\theta_\mathrm{nn}, \theta)$ using the Adam optimiser in which the Glorot uniform was selected to initialise the weight matrix. The network was trained for 30,000 epochs. 

For evaluation purposes, we introduced two metrics. 
The first is the relative squared error for each state variable, $x_{\mathrm{nn},i} (i = 1, \ldots, N)$ as follows:
\begin{align*}
\mathrm{RSE}(x_i) := \frac{\sum_{d=1}^{|\mathcal{T}_\mathrm{val}|}{\left(\bar{x}_i(t_d) - x_{\mathrm{nn},i}(t_d; \hat{\theta}_\mathrm{nn})\right)}^2}{\sum_{d=1}^{|\mathcal{T}_\mathrm{val}|}{{\left(\bar{x}_i(t_d) - \mathrm{Mean}(\bar{x}_i)\right)}^2}}, 
\end{align*}
where $\mathrm{Mean}(\bar{x}_i) = \frac{1}{|\mathcal{T}_\mathrm{val}|}\sum_{d=1}^{|\mathcal{T}_\mathrm{val}|} \bar{x}_i(t_d)$.
This evaluates the accuracy of the state estimation. 
The second is the relative absolute error of each parameter $\hat{\theta}_j (j = 1, \ldots, n)$.  
\begin{align*}
\mathrm{RAE}(\theta_j) := \frac{| \theta_{*,j} - \hat{\theta}_j|}{|\theta_{*,j} |},
\end{align*}
where $\theta_{*,j}$ denotes the $j$th element in $\theta_*$.
The second metric evaluates the accuracy of the parameter estimation.

\subsection{Results}\label{subsec:results}
\subsubsection{Results for Target Scenario 1}
\begin{figure}[htbp]
\centering
\begin{subfigure}[b]{0.48\textwidth}
    \centering
    \includegraphics[width=\linewidth]{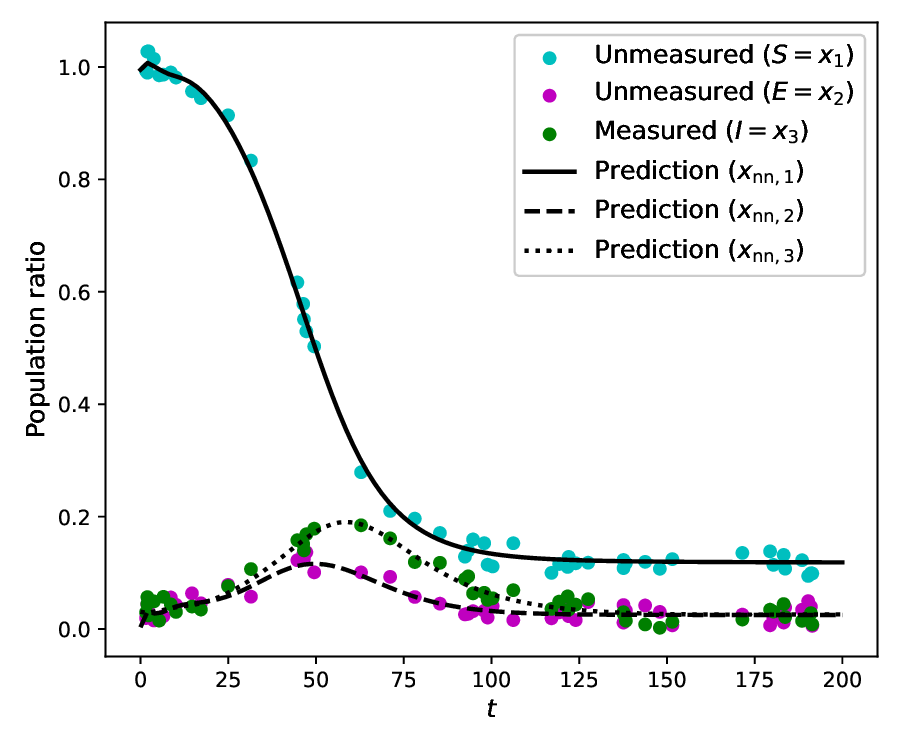}
    \caption{Proposed method.}
    \label{fig:ode1_state_proposed}
\end{subfigure}
\hfill
\begin{subfigure}[b]{0.48\textwidth}
    \centering
    \includegraphics[width=\linewidth]{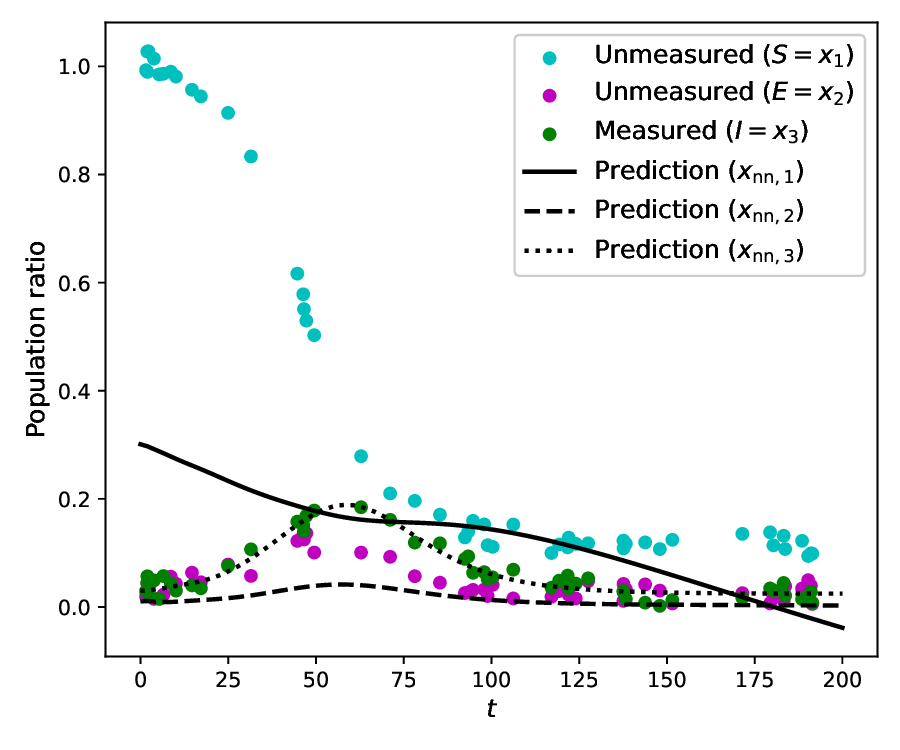}
    \caption{Reference method.}
    \label{fig:ode1_state_reference}
\end{subfigure}
\hfill
\begin{subfigure}[b]{0.48\textwidth}
    \centering
    \includegraphics[width=\linewidth]{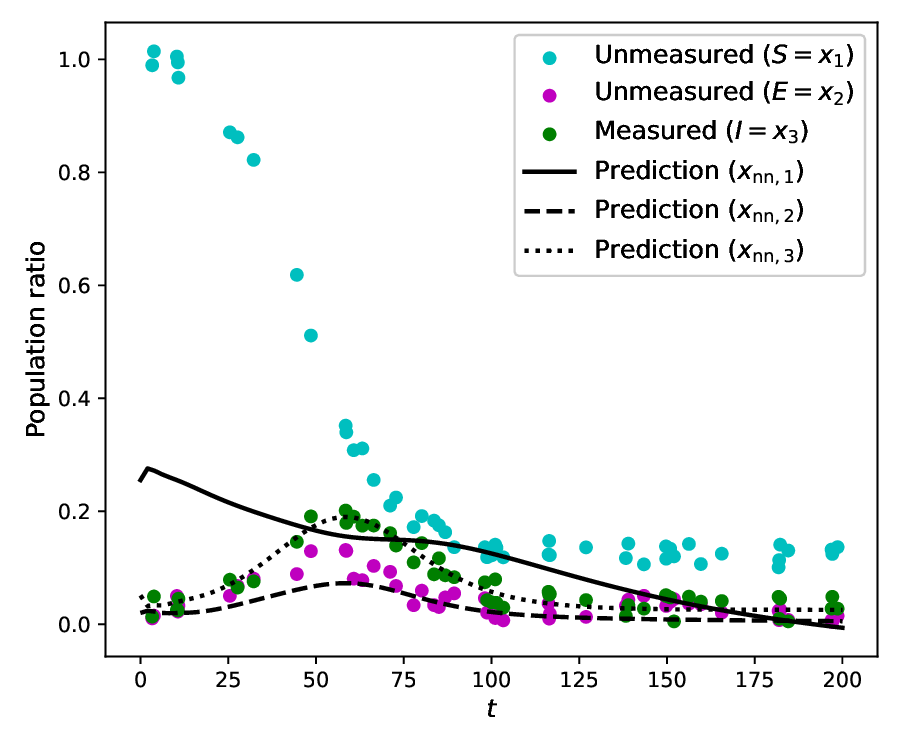}
    \caption{Baseline method.}
    \label{fig:ode1_state_baseline}
\end{subfigure}
\caption{Results of state estimation for Target Scenario 1 given $\sigma = 0.05$ by the proposed, reference and baseline methods given initial condition. Coloured circles represent test data.
The same legend applies to Figs. \ref{fig:ode1_state_reference} and \ref{fig:ode1_state_baseline} as shown in Fig. \ref{fig:ode1_state_proposed}.}
\label{fig:ode1_state}
\end{figure}

Table \ref{tab:rsex_tar1} shows the RSEs of the state variables and RAEs of $\epsilon$ obtained using the proposed, reference, and baseline methods for each noise setting $\sigma = 0.05, 0.1$. 
The RSEs of the measured variable $I = x_3$ remained comparable, irrespective of the method used or the noise level. However, for the unmeasured variables, the proposed methods yield lower RSEs than the other methods for both $\sigma = 0.05, 0.1$. 
For each noise level, the RAEs decreased in the order of reference, baseline, and proposed methods. 
Specifically, the relative absolute error of the values of $\mathrm{RAE}(\epsilon)$ shows that the proposed method achieves an improvement of approximately $83\%$ over the baseline method at $\sigma = 0.05$, and $88\%$ at $\sigma = 0.1$.
Comparing the values of RAEs by the reference and baseline methods, it was suggested that the separation of the training of $\theta_\mathrm{nn}$ and $\epsilon$ has no beneficial effect on the parameter estimation without augmenting the unmeasured data. 
The state estimation results are shown in Fig. \ref{fig:ode1_state}. 
In Fig. \ref{fig:ode1_state}, both the predictions by the neural network after training, $x_\mathrm{nn}(t; \hat{\theta}_\mathrm{nn})$ and test data, where $\sigma = 0.05$, are presented for each method.  
\begin{table}
\caption{Results for Target Scenario 1. RSEs of state variables $(S, E, I, R)$ and RAE of $\epsilon$ obtained using the proposed, reference and baseline methods are shown. \label{tab:rsex_tar1}}
\begin{center}
\begin{tabular}{|c|c|c|c|c|}
\hline
Method & $\mathrm{RSE}(S)$ & $\mathrm{RSE}(E)$ & $\mathrm{RSE}(I)$ & $\mathrm{RAE}(\epsilon)$\\
\hline
Proposed $(\sigma = 0.05)$ & $3.21\times 10^{-2}$ &$2.78\times 10^{-1}$ & $1.66\times 10^{-1}$ & $1.63\times 10^{-2}$ \\
Proposed $(\sigma = 0.1)$  &  $3.08\times10^{-2}$ & $2.28\times10^{-1}$ & $1.41\times10^{-1}$ & $1.24\times 10^{-3}$\\
Reference $(\sigma = 0.05)$ & $7.10\times 10^{-1}$  &  $6.97\times 10^{-1}$  &  $1.76\times 10^{-1}$  & $5.96\times 10^{-1}$\\
Reference $(\sigma = 0.1)$ & $8.94\times10^{-1}$ & $1.69\times10$ & $1.61\times10^{-1}$ & $9.68\times 10^{-1}$ \\ 
Baseline $(\sigma = 0.05)$ & $7.15\times 10^{-1}$ & $ 4.66\times 10^{-1}$ & $ 1.50\times 10^{-1}$ & $9.50\times 10^{-2}$\\
Baseline $(\sigma = 0.1)$ & $7.12\times 10^{-1}$ & $ 6.29\times 10^{-1}$ & $ 2.82\times 10^{-1}$ &$1.00 \times 10^{-2}$ \\
\hline
\end{tabular}
\end{center}
\end{table}

\begin{figure}[htbp]
\centering
\begin{subfigure}[b]{0.49\textwidth}
    \centering
    \includegraphics[width=\linewidth]{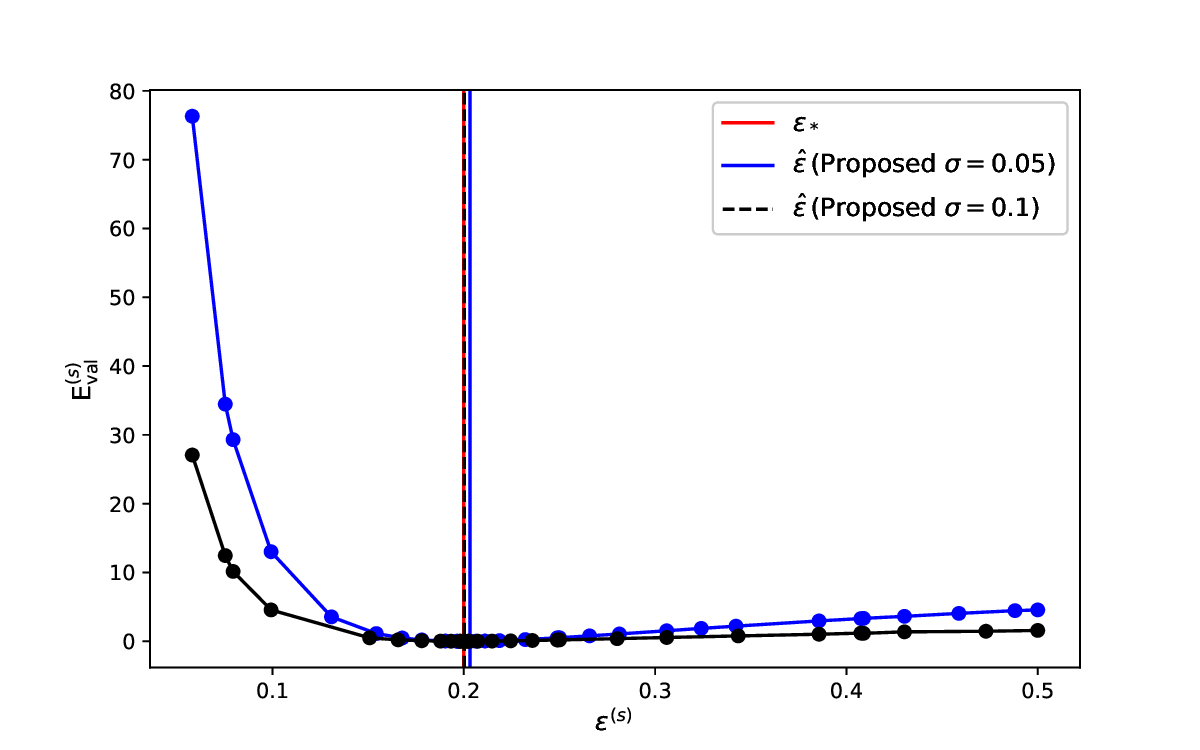}
    \caption{Proposed method.}
    \label{fig:ode1_obj_proposed}
\end{subfigure}
\hfill
\begin{subfigure}[b]{0.49\textwidth}
    \centering
    \includegraphics[width=\linewidth]{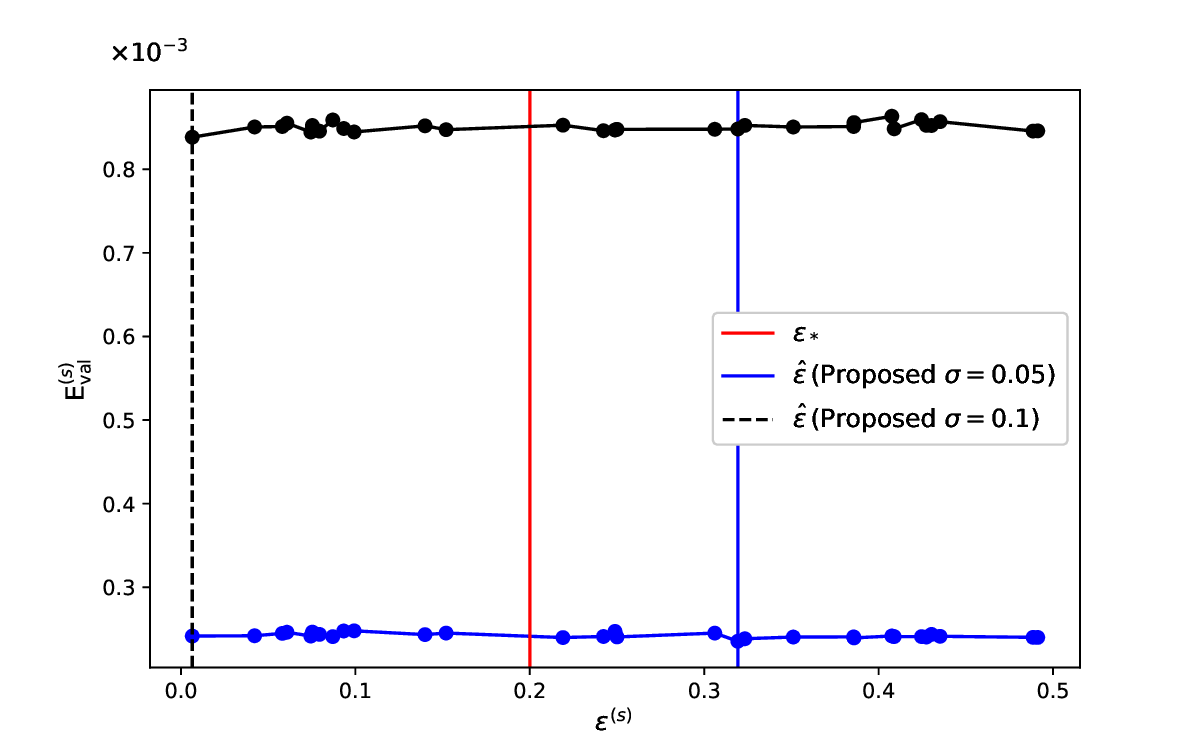}
    \caption{Reference method.}
    \label{fig:ode1_obj_reference}
\end{subfigure}
\caption{Evaluation of \eqref{eq:obj_val}  for Target Scenario 1 obtained by the proposed method and the reference method.}
\label{fig:obj_tar1}
\end{figure}

Fig. \ref{fig:obj_tar1} shows the evaluation of \eqref{eq:obj_val}, the objective function of GP-BO, for the proposed and reference methods, respectively. As can be seen in Fig. \ref{fig:ode1_obj_reference}, the evaluations by the reference method are nearly constant. However, the evaluation of \eqref{eq:obj_val} using the proposed method yields convex-like functions, as shown in Fig. \ref{fig:ode1_obj_proposed}. Furthermore, it can be observed that the curvature increases as the noise decreases, which is reasonable. The near convexity of \eqref{eq:obj_val} demonstrates the positive effects of the separation of the training of $\theta_\mathrm{nn}$ and $\theta$ taking advantage of the augmented data generated based on \eqref{eq:seir_H1} and \eqref{eq:seir_H2}.

\subsubsection{Results for Target Scenario 2}
\begin{figure}[htbp]
\centering
\begin{subfigure}[b]{0.48\textwidth}
    \centering
    \includegraphics[width=\linewidth]{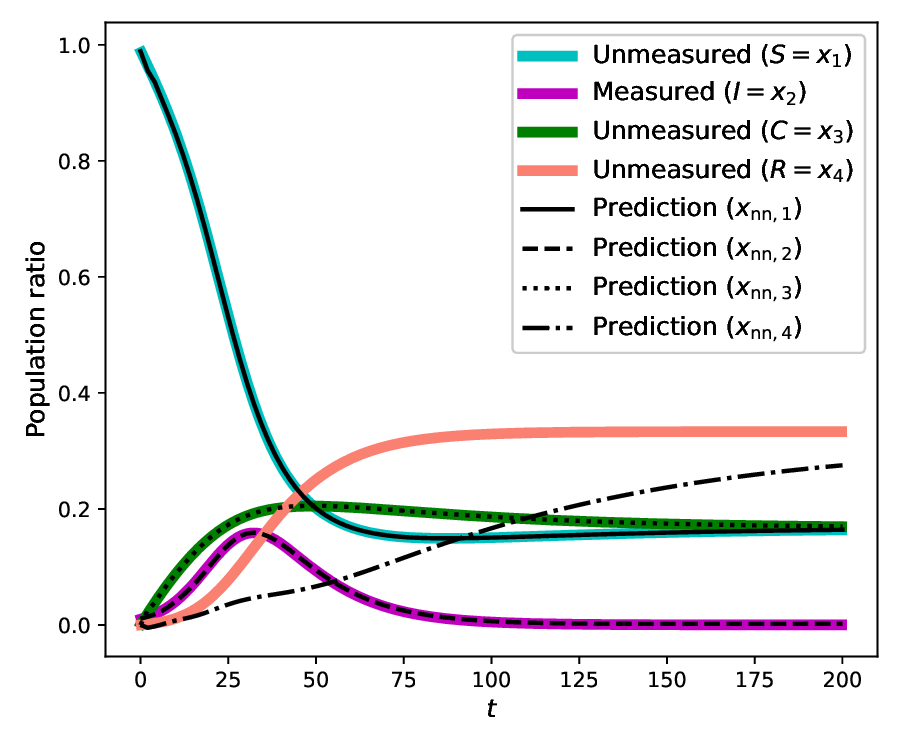}
    \caption{Proposed method.}
    \label{fig:ode2_state_proposed}
\end{subfigure}
\hfill
\begin{subfigure}[b]{0.48\textwidth}
    \centering
    \includegraphics[width=\linewidth]{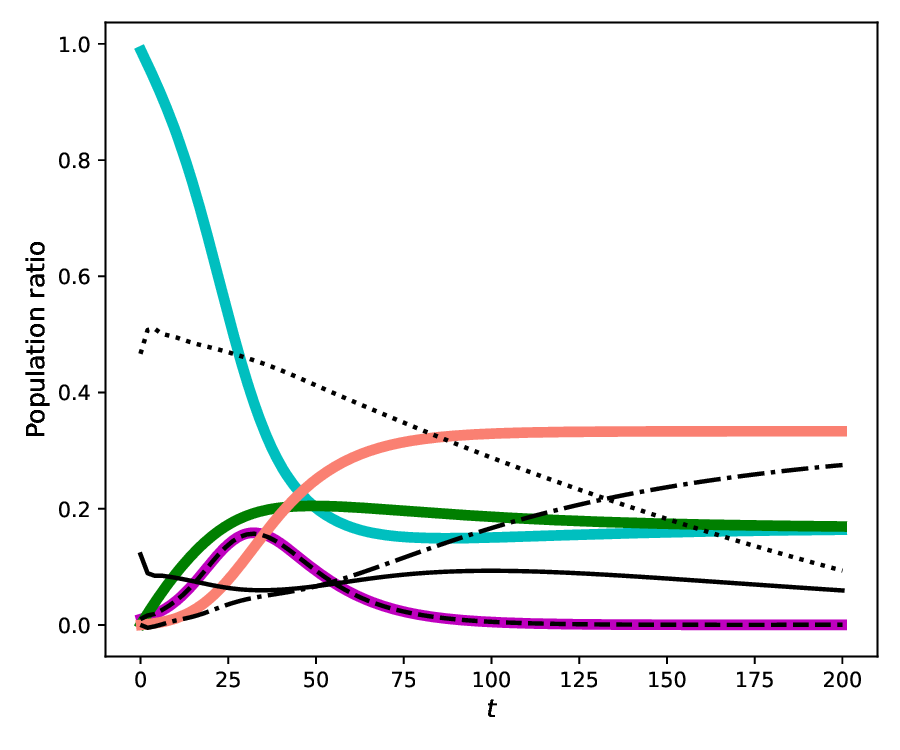}
    \caption{Baseline method.}
    \label{fig:ode2_state_baseline}
\end{subfigure}
\caption{Results of state estimation for Target scenario 2 by the proposed and baseline methods given initial condition of $R = x_4$. 
Coloured and solid lines represent ground truth.
The same legend applies to the right figure as shown in the left.}
\label{fig:ode2_state}
\end{figure}

Table \ref{tab:rsex_tar2} lists the RSEs of the state variables and RAEs of $\beta$ obtained using the proposed and baseline methods. 
The results of state estimation using the proposed and baseline methods, given the initial condition of $R$ are shown in Fig. \ref{fig:ode2_state}. 
As no noise was added to the data used in Target scenario 2, the numerical solutions of \eqref{eq:sicrd}, given $\beta = \beta_*$, were plotted as the ground truth. 
Because of the algebraic unobservablity of $R = x_4$, the values of $\mathrm{RSE}(R)$ obtained using the proposed method are significantly larger than those of the other variables $S, I, C$.  
The RSE values for each variable with and without the initial condition of $R$ showed minimal difference. The $\mathrm{RAE}(\beta)$ obtained with and without $R(0)$ differed by approximately 16$\%$. Although $R$ is algebraically unobservable, the equation $\dot{R} = \gamma I$ is included in \eqref{eq:loss_aug}. Thus, introducing $R(0)$ may restrict the search space of $\beta$. The evaluation of \eqref{eq:obj_val} for the proposed methods with and without the initial condition of $R$ is given by: 
Fig. \ref{fig:ode2_obj}. 
\begin{table}
\caption{Results for Target Scenario 2. RSEs of state variables $(S, I, C, R)$ and RAE of $\epsilon$ obtained using the proposed and baseline methods are shown. \label{tab:rsex_tar2}}
\begin{center}
\begin{tabular}{|c|c|c|c|c|c|}
\hline
Method & $\mathrm{RSE}(S)$ & $\mathrm{RSE}(I)$ & $\mathrm{RSE}(C)$ & $\mathrm{RSE}(R)$ & $\mathrm{RAE}(\beta)$\\
\hline
Proposed (with $R(0)$) & $3.41\times 10^{-3}$ & $2.70\times 10^{-2}$ & $5.74\times 10^{-3}$ & $4.52\times 10^{-1}$ &$7.97\times 10^{-2}$\\
Proposed (without $R(0)$)& $2.33\times 10^{-3}$ & $2.92\times 10^{-2}$ & $5.62\times 10^{-3}$ & $3.60\times 10^{-1}$ & $2.46\times 10^{-1}$\\
Baseline (with $R(0)$)& $8.24\times 10^{-1}$ & $1.03\times 10^{-2}$ & $1.02$ & $2.33\times 10^{-1}$ & $2.75$\\
\hline
\end{tabular}
\end{center}
\end{table}
\begin{figure}[htbp]
\centering
\begin{subfigure}[b]{0.6\textwidth}
    \centering
    \includegraphics[width=\linewidth, height=5cm]{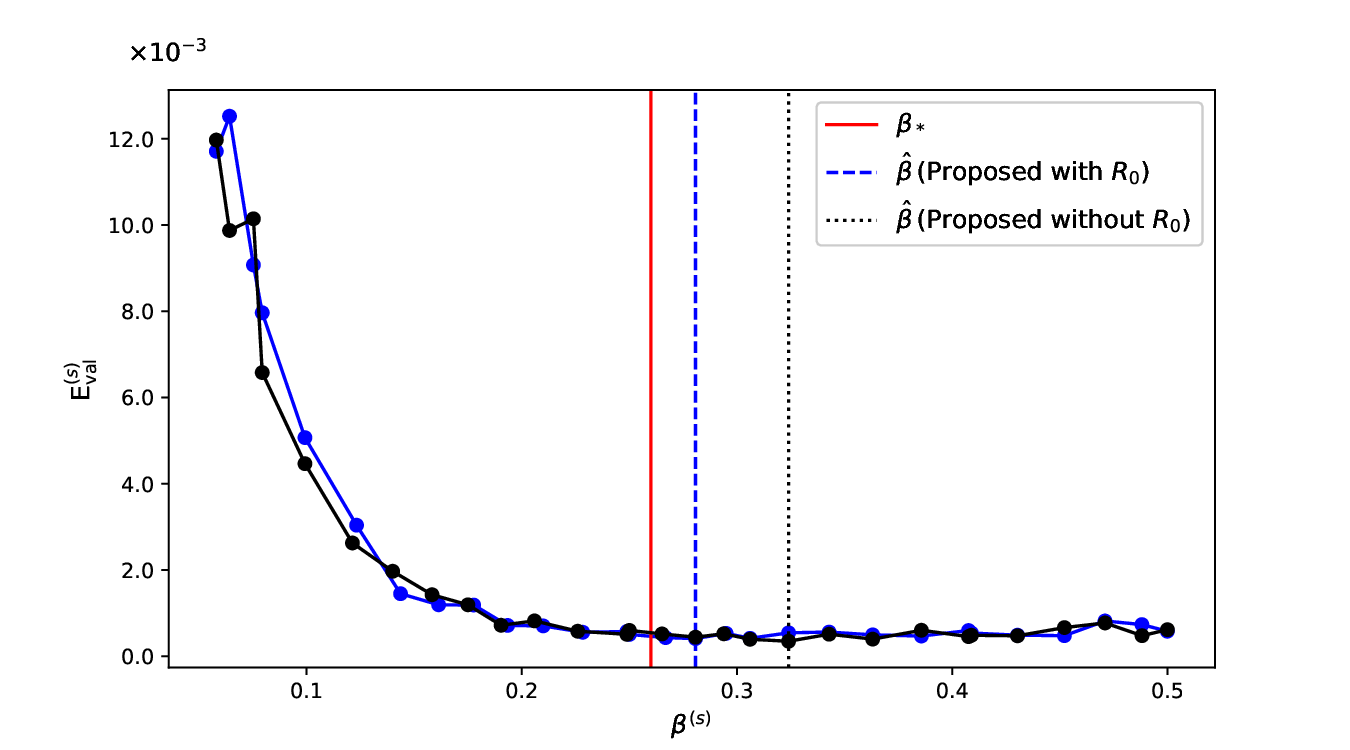}
    \caption{Evaluations of \eqref{eq:obj_val} by the proposed method.}
    \label{fig:ode2_obj}
\end{subfigure}
\hfill
\begin{subfigure}[b]{0.38\textwidth}
    \centering
    \includegraphics[width=\linewidth, height=5cm]{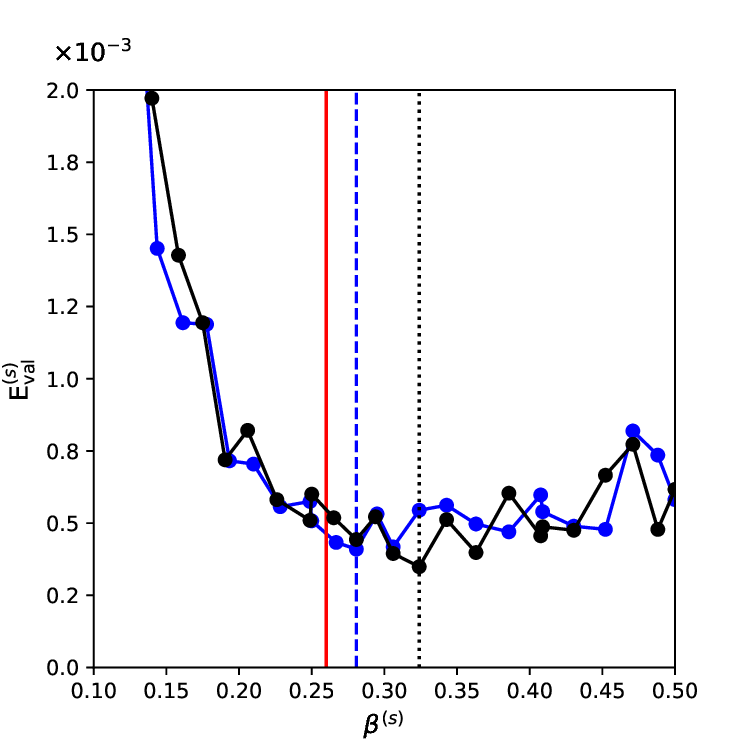}
    \caption{Zoomed-in version of Fig. \ref{fig:ode2_obj}.}
    \label{fig:ode2_obj_zoom}
\end{subfigure}
\caption{Evaluation of \eqref{eq:obj_val} for Target Scenario 2 obtained by the proposed method (with and without the initial condition of $R$). On Fig. \ref{fig:ode2_obj}(a), the evaluation is plotted without zooming in, and on Fig. \ref{fig:ode2_obj}(b), a zoomed-in version of the same plot is presented. The same legend applies to Fig. \ref{fig:ode2_obj}(b) as shown in Fig. \ref{fig:ode2_obj}(a).}
\label{fig:ode2_obj}
\end{figure}

\subsubsection{Results for Target Scenario 3}
\begin{figure}[htbp]
\centering
\begin{subfigure}[b]{0.48\textwidth}
    \centering
    \includegraphics[width=\linewidth]{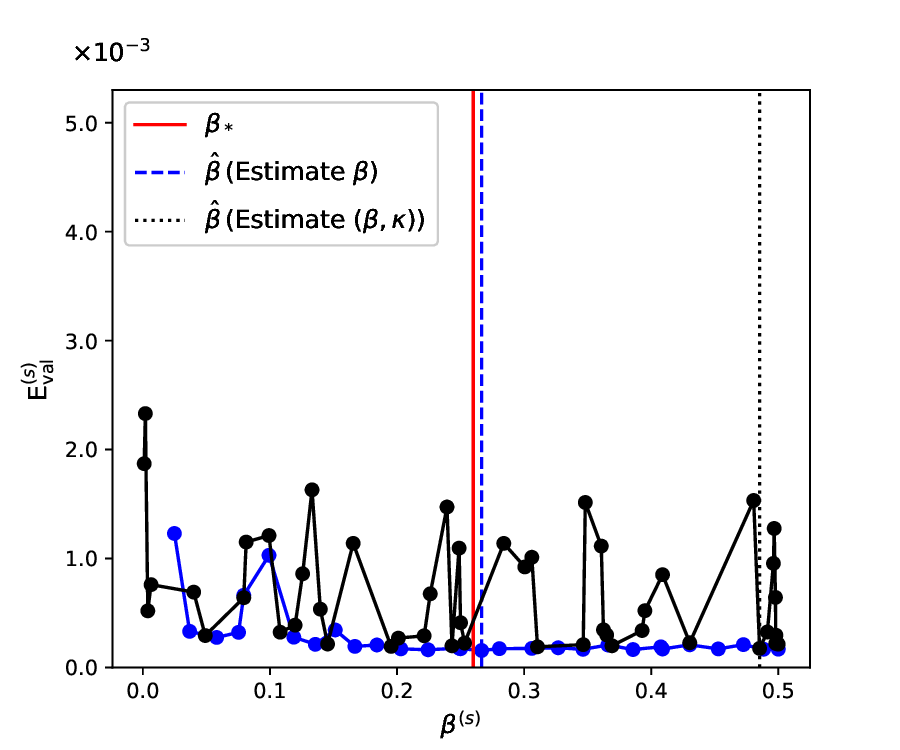}
    \caption{Evaluation of \eqref{eq:obj_val} projected onto the $\beta$-axis.}
    \label{fig:ode3_beta_obj}
\end{subfigure}
\hfill
\begin{subfigure}[b]{0.48\textwidth}
    \centering
    \includegraphics[width=\linewidth]{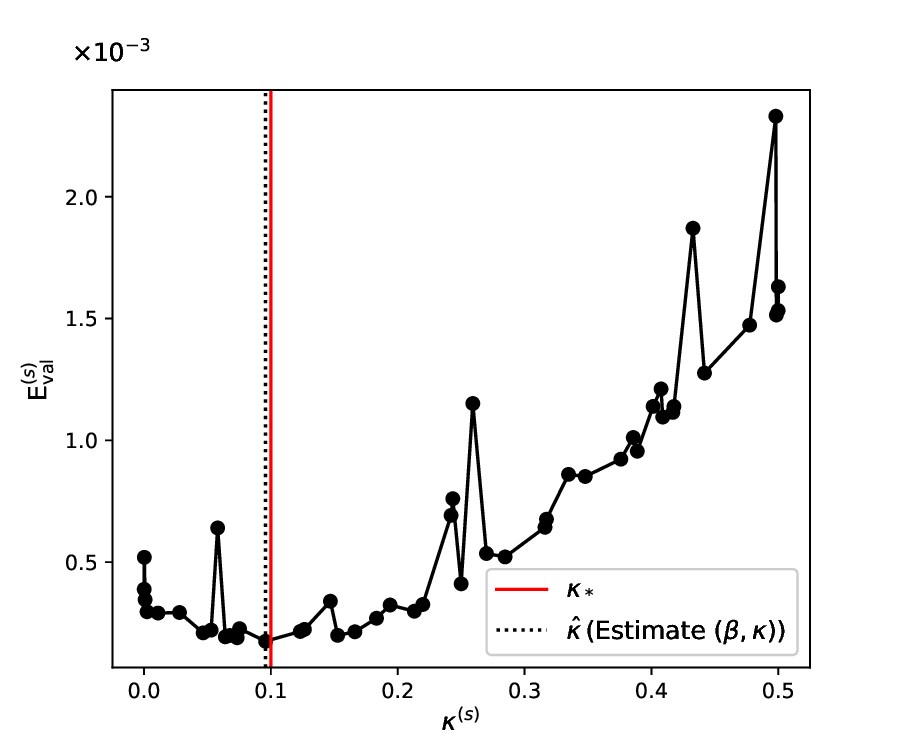}
    \caption{Evaluation of \eqref{eq:obj_val} projected onto the $\kappa$-axis.}
    \label{fig:ode3_kappa_obj}
\end{subfigure}
\caption{Evaluation of \eqref{eq:obj_val} for Target Scenario 3 obtained by the proposed method. 
In Fig. \ref{fig:ode3_beta_obj}, the blue lines show the result where $\beta$ is the unknown parameter. In Figs. \ref{fig:ode3_beta_obj} and \ref{fig:ode3_kappa_obj}, the black lines show the results where $(\beta, \kappa)$ is the unknown parameter vector.  
}
\label{fig:ode3_obj}
\end{figure}

Table \ref{tab:rsex_tar3} lists the RSEs of the state variables and the RAEs of $\beta, \kappa$ obtained using the proposed methods. The first row corresponds to the results in which both $\beta$ and $\kappa$ are unknown. The second row corresponds to the results, where $\beta$ is unknown and $\kappa$ is fixed to $\kappa_*$.
Fig. \ref{fig:ode3_obj} shows the evaluation of \eqref{eq:obj_val} for the proposed method. 
As can be seen from the second row of Table \ref{tab:rsex_tar3}, the generally small RSE and RAE values show the applicability of the proposed method, even when an input is introduced into the epidemiological model. The values of RSEs between the first and second rows show minimal difference; however, $\mathrm{RAE}(\beta)$ differs by $84\%$. This demonstrates the difficulty of learning the parameter vector $\theta \in \mathbb{R}^n$, where $n > 1$. 
Notably, $\mathrm{RAE}(\kappa)$ is sufficiently small, as the first row shows, where $\mathrm{RAE}(\beta)$ is 0.866. This suggests the necessity of improving sampling strategies to account for differences in parameter sensitivity.
\begin{table}
\caption{Results for Target Scenario 3. RSEs of state variables $(S, A, I, R)$ and RAE of $(\beta, \kappa)$ obtained using the proposed and baseline methods are shown. \label{tab:rsex_tar3}}
\begin{center}
\begin{tabular}{|c|c|c|c|c|c|c|c|}
\hline
Method & $\mathrm{RSE}(S)$ & $\mathrm{RSE}(A)$ & $\mathrm{RSE}(I)$  & $\mathrm{RAE}(\beta)$ & $\mathrm{RAE}(\kappa)$ \\
\hline
Proposed $(\theta = (\beta, \kappa))$ & $2.06\times 10^{-3}$ & $4.28\times 10^{-3}$ & $4.30\times 10^{-2}$ & $8.66\times 10^{-1}$ & $ 4.18\times 10^{-2}$\\
Proposed $(\theta = \beta)$ &  $1.59\times10^{-3}$ & $9.02\times10^{-3}$ & $1.43\times10^{-2}$ & $2.53\times 10^{-2}$ &  $-$\\
\hline
\end{tabular}
\end{center}
\end{table}

\section{Discussion and Conclusion}\label{sec:conclusion}
This study proposes an algebraically observable PINN that can perform state estimation and parameter estimation given partial measurements.
Algebraically observable PINN exploits the algebraic observability of unmeasured variables, which has been ignored in previous studies. 
The unmeasured data were augmented if the corresponding variable was algebraically observable. This enables a more accurate state and parameter estimation compared with conventional methods and does not require careful hyperparameter tuning of the weighting constants that appear in the loss function.
The key technique of an algebraic observable PINN is the algebraic observability analysis, which can be automatically conducted using computer algebra software. 

The validity of the proposed method was demonstrated using three scenarios, each of which considered a distinct epidemiological model as the governing equation for the measured data. 
Owing to data augmentation based on algebraic observability, the proposed method achieves better accuracy than the baseline method in the estimation of unmeasured variables across Target Scenarios 1 and 2.
Through numerical experiments for Target Scenario 1, the robustness of the parameter estimation using the proposed method was numerically demonstrated. In addition, the separation of the neural network and epidemiological parameters during training and data augmentation based on the results of the algebraic observability analysis was confirmed to be important. 
In numerical experiments for Target Scenarios 2 and 3, the effectiveness of the proposed method was further demonstrated in more advanced problem settings, where the input function and algebraically unobservable variables were present in epidemiological models. 

We conclude this paper with a discussion of future work. 
Although not explored in depth in this study, incorporating techniques such as \cite{AAAalg} and \cite{sobolev} to reduce numerical errors in estimating higher-order derivatives of data remains an important issue for practical applications. 
Another key challenge is the refinement of the parameter sampling strategy. 
Although GP-BO was employed in this study, more advanced methods that account for parameter sensitivity may be necessary. Such improvements are expected to enhance the estimation accuracy when multiple unknown parameters are involved and reduce the overall computational cost. This has also been explored by using methods based on particle filters \cite{comp_pf}, and incorporating insights from these studies may be beneficial. 
A detailed comparison between the proposed method and these approaches, as well as the potential integration of these methods, remains for future work.

\begin{credits}
\subsubsection{\ackname} This work is supported by JST ACT-X Grant JPMJAX22A7, JSPS KAKENHI Grand Number 22K21278 and 24K16963.

\subsubsection{\discintname}
The authors declare no conflicts of interest.
\end{credits}

\appendix
\section[]{Appendix}\label{sec:singular_ex}
The following is an example of a series of commands for Singular \cite{singular} used to investigate the algebraic observability shown in Example \ref{example:seir_ao_ex}. See Table \ref{table:ex1x2} for the notations of the variables and parameters of \eqref{eq:seir} in the commands. See the Appendix \cite{robot} for details on the commands.

\begin{lstlisting}[frame=lines, breaklines=true,caption=Singular commands to check the algebraic observability of $x_2$ of \eqref{eq:seir_x}, label=listing:ex1x2]
ring r =(0, theta1, theta2, theta3),(x1,x3,d1x1,d1x2,d1x3,d2x1,d2x2,d2x3,x2, y, d1y, d2y),(lp(8),lp(1),lp(3));
ideal J1 = d1x1 + theta1*x1*x3, d2x1 + theta1*d1x1*x3 + theta1*x1*d1x3, d1x2 - theta1*x1*x3 + theta2*x2, d2x2 - theta1*d1x2*theta2 - theta1*x1*d1x3 + theta2*d1x2, d1x3 - theta2*x2 + theta3*x3, d2x3 - theta2*d1x2 + theta3*d1x3, y - x3, d1y - d1x3,d2y - d2x3;
option(redSB);
groebner(J1);
\end{lstlisting}

\begin{lstlisting}[frame=lines, breaklines=true,caption=Output of Listing \ref{listing:ex1x2}, label=listing:ex1x2out]
(theta2)*x2+(-theta3)*y-d1y,
d2x3-d2y,
(theta2)*d2x2*y+(-theta1*theta2*theta3)*y*d1y+(-theta1*theta2+theta2)*y*d2y+(-theta2-theta3)*d1y*d1y-d1y*d2y,
(theta2)*d2x1+(theta2)*d2x2+(-theta1*theta2*theta3)*y*y+(-theta1*theta2-theta1*theta3)*y*d1y+(-theta1)*y*d2y+(-theta1*theta2*theta3+theta2*theta3)*d1y+(-theta1*theta2+theta2)*d2y,
d1x3-d1y,
(theta2)*d1x2+(-theta3)*d1y-d2y,
(theta2)*d1x1+(theta2*theta3)*y+(theta2+theta3)*d1y+d2y,
x3-y,
(theta1)*x1*d1y-d2x2+(theta1*theta3-theta3)*d1y+(theta1-1)*d2y,
(theta1*theta2)*x1*y+(-theta2*theta3)*y+(-theta2-theta3)*d1y-d2y
\end{lstlisting}

\begin{table}
\caption{Correspondence of variables of Listing \ref{listing:ex1x2}, \ref{listing:ex1x2out}, \eqref{eq:seir_x}, and \eqref{eq:seir}. \label{table:ex1x2}}
\begin{center}
\begin{tabular}{|c|c|c|}
\hline
Listing \ref{listing:ex1x2}, \ref{listing:ex1x2out} & \eqref{eq:seir_x}, \eqref{eq:seir_measure} & \eqref{eq:seir}\\
\hline
(x1, x2, x3)		& $(x_1, x_2, x_3)$ & $ (S, E, I)$  \\
(d1x1, d1x2, d1x3)   & $(\dot{x}_1, \dot{x}_2, \dot{x}_3) $ & $(\dot{S}, \dot{E}, \dot{I})$\\
(d2x1, d2x2, d2x3)   & $(\ddot{x}_1, \ddot{x}_2, \ddot{x}_3)$ & $ (\ddot{S}, \ddot{E}, \ddot{I})$\\
(y, d1y, d2y)        & $(y,\dot{y}, \ddot{y})$ & \\
(theta1, theta2, theta3)	& $(\beta, \gamma, \epsilon) $ & $ (\beta, \epsilon, \gamma)$	\\
\hline
\end{tabular}
\end{center}
\end{table}

\begin{lstlisting}[frame=lines, breaklines=true,caption=Specification of the ring to check the observability of $x_1$ in \eqref{eq:seir_x}, label=listing:ex1x1]
ring r =(0, theta1, theta2, theta3),(x2,x3,d1x1,d1x2,d1x3,d2x1,d2x2,d2x3, x1, y, d1y, d2y),(lp(8),lp(1),lp(3));
\end{lstlisting}

\begin{lstlisting}[frame=lines, breaklines=true,caption=Output of Listing \ref{listing:ex1x2} with the ring defined in Listing \ref{listing:ex1x1}, label=listing:ex1x1out]
d2x3-d2y,
d2x2+(-theta1)*x1*d1y+(-theta1*theta3+theta3)*d1y+(-theta1+1)*d2y,
(theta2)*d2x1+(theta1*theta2)*x1*d1y+(-theta1*theta2*theta3)*y*y+(-theta1*theta2-theta1*theta3)*y*d1y+(-theta1)*y*d2y,
d1x3-d1y,
(theta2)*d1x2+(-theta3)*d1y-d2y,
(theta2)*d1x1+(theta2*theta3)*y+(theta2+theta3)*d1y+d2y,
x3-y,
(theta2)*x2+(-theta3)*y-d1y
\end{lstlisting}

%
%
%

\bibliographystyle{splncs04}
\bibliography{ref}
\end{document}